\documentclass[apjl,twocolumn]{openjournal}
\usepackage{amsmath}
\usepackage{booktabs}
\usepackage{multirow}
\usepackage{color}
\usepackage{soul}
\usepackage{upgreek}
\usepackage{natbib}

\usepackage[dvipsnames]{xcolor} %added later for color

\usepackage{hyperref}
\hypersetup{
    colorlinks=true,
    linkcolor=blue,
    breaklinks=true,      
    urlcolor=blue,
    citecolor=blue,
}
\usepackage{orcidlink}

\renewcommand{\arraystretch}{1.2}  
\usepackage{listings}
\usepackage{color}
\definecolor{dkgreen}{rgb}{0,0.6,0}
\definecolor{gray}{rgb}{0.5,0.5,0.5}
\definecolor{mauve}{rgb}{0.58,0,0.82}
\definecolor{golden}{rgb}{0.86,0.65,0.01}
\newcommand{\unitHa}{$\; \rm ergs \, \rm s^{-1} \, \rm kpc^{-2}$}
\newcommand{\msun}{$\; \rm \mathrm{M}_\odot$}

\begin{document}

\title{Star formation beyond galaxies: widespread in-situ formation of intra-cluster stars \vspace{-1.5cm}}

\author{
Niusha Ahvazi,$^{1,2,3,\star} \orcidlink{0009-0002-1233-2013}$%
Laura V. Sales,$^{1} \orcidlink{0000-0002-3790-720X}$
Julio F. Navarro, $^{4} \orcidlink{0000-0003-3862-5076}$
Andrew Benson,$^{2} \orcidlink{0000-0001-5501-6008}$
Alessandro Boselli,$^{5}$
and Richard D'Souza$^{6} \orcidlink{0000-0001-9269-8167}$
\\
% List of institutions
$^{1}$Department of Physics and Astronomy, University of California, Riverside, 900 University Avenue, Riverside, CA 92521, USA\\
$^{2}$Carnegie Observatories, 813 Santa Barbara Street, Pasadena, CA 91101, USA\\
$^{3}$Department of Astronomy, University of Virginia, 530 McCormick Road, Charlottesville, VA 22904, USA\\
$^{4}$Department of Physics and Astronomy, University of Victoria, Victoria BC V8W 3P6, Canada\\
$^{5}$Aix Marseille Univ, CNRS, CNES, LAM, Marseille, France\\
$^{6}$Vatican Observatory, Specola Vaticana, V-00120, Vatican City State
}

\thanks{$^\star$E-mail:} \email{niuhsa.ahvazi@email.ucr.edu}
\email{nahvazi@carnegiescience.edu}
\email{nahvazi@virginia.edu}

\begin{abstract}
We study the fraction of the intra-cluster light (ICL) formed in-situ in the three most massive clusters of the TNG50 simulation,  with virial masses $\sim 10^{14}$\msun. We find that a significant fraction of ICL stars ($8\%$-$28\%$) are born in-situ. This amounts to a total stellar mass comparable to the central galaxy itself. Contrary to simple expectations, only a sub-dominant fraction of these in-situ ICL stars are born in the central regions and later re-distributed to more energetic orbits during mergers. Instead, many in-situ ICL stars form directly hundreds of kiloparsecs away from the central galaxy, in clouds condensing out of the circum-cluster medium.\ The simulations predict a present-date diffuse star formation rate of $\sim$1 $\mathrm{M}_{\odot}$/yr, with higher rates at higher redshifts. The diffuse star forming component of the ICL is filamentary in nature, extends for hundreds of kiloparsecs and traces the distribution of neutral gas in the cluster host halo.\ We discuss briefly how numerical details of the baryonic treatment in the simulation, in particular the density threshold for star formation and the equation of state, may play a role in this result. We conclude that a sensitivity of $1.6 \times 10^{-19} - 2.6 \times 10^{-18}$ erg s$^{-1}$ cm$^{-2}$ arcsec$^{-2}$ in H$_\alpha$ flux (beyond current observational capabilities) would be necessary to detect this diffuse star-forming component in galaxy clusters. 
\keywords{galaxies: clusters: intracluster medium -- Galaxy: formation -- Galaxy: evolution}

\end{abstract}

\maketitle  

\section{Introduction}

The simplest models of galaxy formation suggest that most stars in a cluster should form in the high-density regions of individual galaxies. These models imply that the majority of stars in the intra-cluster light (ICL) originate from the tidal removal of stars from luminous galaxies with mass similar to the Milky Way (MW) \citep{Puchwein_2010, Cui2014, Cooper2015, Contini2013, Contini2018, Contini2019, Montenegro2023}, with relatively small contributions from  dwarf mass galaxies, particularly in group environments with halo virial masses $<10^{13}\; \rm \mathrm{M}_\odot$ \citep{Ahvazi2023}. Nevertheless,  observational studies \citep{Barfety2022, Hlavacek2020} and some numerical simulations \citep{Puchwein_2010} indicate that a non-negligible fraction of ICL stars are formed in-situ.

In-situ ICL stars could have formed in the brightest cluster galaxy (BCG) galaxy and been later pushed onto highly energetic orbits  by mergers. An alternative, more intriguing, possibility is that stars form directly in the intra-cluster medium (ICM), far away from the high density regions of the BCG. This low-density mode of star formation is potentially interesting, as it might occur under very different physical conditions than the normal star formation in galaxies, with  consequences for the multi-phase nature of the ICM and its metallicity, as well as for the age distribution of  ICL stars, especially if in-situ star formation continues to the present day. From an observational perspective, if a substantial amount of stars form outside galaxies, this would have strong implications for wide-field surveys in the UV or H$_\alpha$, which are especially sensitive to the presence of young stars.

Several mechanisms have been suggested to explain in-situ ICL star formation. One of the most commonly cited is the compression of gas associated with the ram-pressure trails of infalling galaxies, a phenomenon that has been detected in both observations \citep{Smith2010, Mihos_2016, Boselli2018, Gullieuszik2020} and some numerical simulations (\citealt{Kronberger2008, Kapferer2009, Tonnesen2012, Lee2020, Calura2020, Steyrleithner2020, Vollmer2021, Goller2023}, for a detailed discussion see section 5.3.1 in review paper by \citealt{Boselli2022}). Another possible mechanism is star formation within the tails of stripped material formed during gravitational fly-by encounters of gas-rich systems (see see NGC 4254 in Virgo, \citealt{Boselli2018VESTIGEIII}; VCC 1249, \citealt{ArrigoniBattaia2012}; IC3418, \citealt{Hester2010, Fumagalli2011, Kenney2014}).

In-situ star formation in the ICM is also predicted to be widespread at high-redshift,  where AGN feedback might be less powerful and unable to offset the short timescales for cooling in the dense regions of the ICM \citep{Webb2015b}. In such a scenario, spontaneous run-away cooling in the ICM might lead to an increase of density high enough to trigger violent star-formation events in locations outside galaxies \citep{Hlavacek2020}. 

Interestingly, observations of off-center star formation at large distances from the BCG are consistent with this scenario for a few high-$z$ clusters \citep{Webb2015, Hlavacek2020, Barfety2022}, in some cases predicting as much as a MW-worth of stars contributed to the ICM by such cooling events. 

Some observations of isolated HII regions in nearby clusters, such as the Virgo cluster, may also hint at a more sustained but lower-level star formation activity taking place directly in the ICM/ICL region \citep{Gerhard2002, Bellazzini2018}. However, it is not always trivial to disentangle whether these isolated star-forming clouds are ICM-born material or belong to some older ram-pressure or interaction events in the cluster \citep{Junais2021, Jones2022}. 

%%%%%%%%%%%%%%%%%%%%%%%%%%%%%%%%%%%%%%%%%%%%%%%%%%
\begin{figure*}
	\includegraphics[width=2.09\columnwidth]{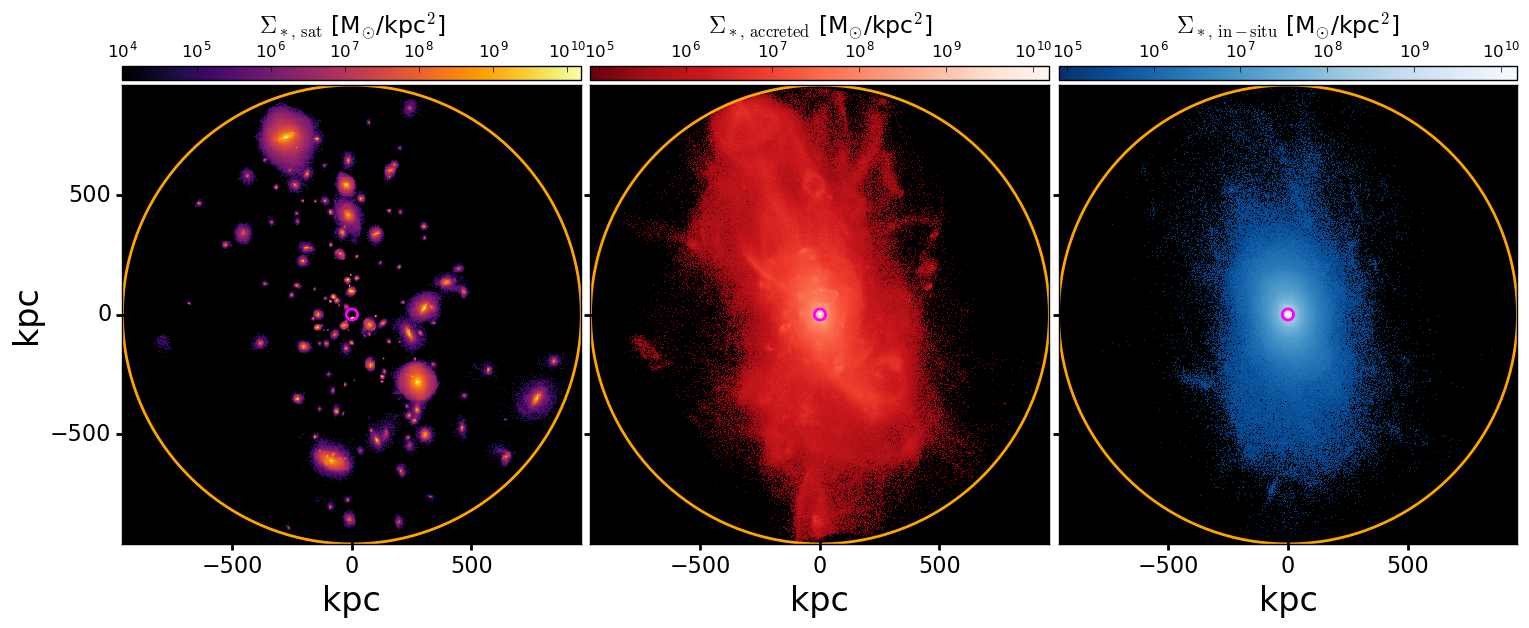}
    \caption{Projected density maps for different stellar components in Group 1 at $z=0$: bound satellites (left), accreted ICL (middle), and in-situ formed ICL (right). The magenta and orange circles indicate twice the stellar half-mass radius of the stars in the central host galaxy ($2 \, \rm r_{h_{*}}$) and the virial radius ($\rm r_{200}$), respectively. These are the bounds used to define the ICL component in our study.}
    \label{fig:sat_stellar}
\end{figure*}
%%%%%%%%%%%%%%%%%%%%%%%%%%%%%%%%%%%%%%%%%%%%%%%%%%

Numerical simulations therefore emerge as valuable tools to explore this mode of star formation within the realistic framework of the $\Lambda$CDM model. For instance, \citet{Puchwein_2010} estimate that approximately $\sim 30\%$ of the ICL stars formed in their clusters (with masses between $2 \times 10^{13}$ and $1.5 \times 10^{15} \, \rm \mathrm{M}_\odot$) originated at distances consistent with being ICM-born. The presence of in-situ formed stars in extended stellar halos is also documented extensively in the literature on the scale of MW-mass halos \citep{Zolotov2009, Font2011, Tissera2012, Cooper2015_2, Pillepich2015}. %For example, the Illustris simulations predict a relatively large fraction of in situ stars at large galacto-centric distances beyond 30 kpc \citep{Pillepich2014} although these predictions appear incompatible with some HST observations (see \citet{DSouza2018}).}

Various numerical methods have been employed to study the formation of cold gas in the ICM, which could give rise to  star formation outside galaxies. Runaway local production of cold gas in the ICM, as well as thermal instability, have been shown to lead to  multi-phase structures in idealized hot haloes \citep{Maller2004, Kaufmann2006, Kaufmann2009, McCourt2012, Sharma2012}. 

Given that the cold-phase ICM resulting from such processes is a small-scale phenomenon, it has been studied largely in idealized numerical experiments \citep{Fielding2020}. Previous work by \cite{Oppenheimer2018} demonstrated a relatively significant abundance of cold, low-ionization gas in the circum-galactic medium (CGM) of EAGLE galaxies (for MW and group mass halos), albeit with uncertainties regarding the numerical robustness of the hydrodynamical technique in this regime (see also \citealt{Ford2013}). Similarly, \cite{Butsky2019} demonstrated evidence for the presence of cold gas in the outskirts of the galaxy clusters (in the ICM) using the high-resolution hydrodynamical galaxy cluster simulation, RomulusC.

Semi-analytic models have also been developed to explain the unique, possibly drag-induced sub-virial kinematics of cold gas flowing through luminous red galaxy host haloes \citep{Afruni2019}. In the context of IllustrisTNG, \cite{Nelson2018} showed that TNG is consistent with empirical OVI constraints in the CGM of MW and group mass halos. 

The cold gas content of galaxies themselves (both neutral HI and molecular H2) has also been favorably compared against data in several regimes, although simplifying assumptions are needed to derive these cold-phase fractions which are not self-consistently followed in typical galaxy-scale simulations. \citet{Nelson2020} also demonstrated that group mass haloes hosting massive galaxies are filled with a hot, virialized plasma at $\sim 10^7$ K, and that they contain a large abundance of cold gas with a characteristic temperature of $\sim 10^4$ K. This cold phase of the ICM takes the form of thousands of small, $\sim$ kpc-sized clouds, which fill the inner halo out to hundreds of kpc (similar to predictions by \citealt{Maller2004} for MW-mass halos). 

There are, however, several important caveats when studying the formation of star-forming clouds in the ICM, mostly connected with limitations of numerical techniques. Approaches such as SPH have been shown to be unable to track the evolution of instabilities, such as the Kelvin-Helmholtz instability, leading to the formation of more dense clouds that are artificially long-lived compared to similar numerical experiments run with grid-base techniques \citep{Agertz2009, Torrey2012}. 

The lack of, or inadequate treatment of other important physics, such as heat conduction, magnetic fields or metal diffusion might also complicate the interpretation of results from simulations (see the discussion in \citealt{Puchwein_2010}). It is clear that, for any numerical technique chosen, high resolution becomes key to study  in-situ star formation in low-mass clouds embedded in the hot ICM of clusters, especially considering the low masses of these clouds. 

In this context, we present here a study of the in-situ formation of stars and their contribution to the ICL using the three most massive groups/clusters in the TNG50 simulations \citep{Pillepich2019,Nelson2019_2}%{\citep{2018MNRAS.475..648P, Nelson2019}}
, one of the highest-resolution cosmological simulations of low-mass clusters with virial mass $\sim 10^{14}\; \rm \mathrm{M}_\odot$ at $z=0$. 

This article is organized as follows: in Section~\ref{method}, we provide a comprehensive overview of the simulation employed, including some definitions crucial for the subsequent sections. Section~\ref{result} presents our findings, together with an in-depth discussion. Finally, Section~\ref{conclusions} summarizes the key conclusions drawn from this study.

\section{Simulation}\label{method}

We use the cosmological hydrodynamical simulation IllustrisTNG \citep{2018MNRAS.475..648P, 2018MNRAS.475..676S, 2018MNRAS.480.5113M, 2018MNRAS.477.1206N, Nelson2018_2, Nelson2019}%{\citep{2018MNRAS.475..648P, 2018MNRAS.475..676S, 2018MNRAS.480.5113M, 2018MNRAS.477.1206N, Nelson2019}}
. The IllustrisTNG project\textsuperscript{1}\footnote{\url{https://www.tng-project.org}} is a suite of state-of-the-art cosmological galaxy formation simulations that come in different physical sizes and mass resolutions. In this work we use the highest resolution baryonic run TNG50-1  (TNG50 for short) \citep{Pillepich2019,Nelson2019_2}, which has a volume of approximately $50 ^{3} \, \rm Mpc^{3}$. It has  $2 \times 2160^3$ resolution elements, and the baryon and dark matter particle masses are $8.5 \times 10^4 \mathrm{M}_{\odot}$ and $4.5 \times 10^5 \mathrm{M}_{\odot}$, respectively. The IllustrisTNG cosmological parameters are consistent with $\Lambda$CDM model determined by Planck XIII \citep{2016A&A...594A..13P} to be $\Omega_{m} = 0.3089,\, \Omega_{b} = 0.0486,\, \Omega_{\Lambda} = 0.6911,\, H_{0} = 100 \, h \mathrm{km s}^{-1}\mathrm{Mpc}^{-1}$ with $h = 0.6774,\, \sigma_{8} = 0.8159,$ and $n_\mathrm{s} = 0.9667.$

Gravity and baryons are followed using the {\sc Arepo} moving mesh code \citep{2010MNRAS.401..791S}. The baryonic treatment is based on the code used for its predecessor simulation suite, Illustris \citep{2013MNRAS.436.3031V, 2014MNRAS.444.1518V, 2014Natur.509..177V, 2014MNRAS.445..175G, 2015A&C....13...12N}, with modifications implemented to better track the formation and evolution of galaxies (as described in \citealp{2018MNRAS.473.4077P, Weinberger2017}).

The updated IllustrisTNG sub-grid models account for star formation, radiative metal cooling, chemical enrichment from SNII, SNIa, and AGB stars, stellar feedback, and super-massive black hole feedback \citep{Weinberger2017, 2018MNRAS.473.4077P}. The star formation in the Interstellar Medium (ISM) is treated in a sub-resolution fashion following a variation of the \cite{Springel2003} with the adoption of a \cite{Chabrier2003} initial mass function, and model the star-forming dense ISM gas using an effective equation of state where stars form stochastically above a gas density of $\rho_{\rm sfr}$ = $0.13$ $cm^{-3}$ with a star formation time-scale of $t_{\rm sfr}$ = $2.2$ Gyr (refer to \citealt{Vogelsberger2013} for detailed explanations). These models were calibrated to reproduce a set of observational constraints that include the observed $z = 0$ galaxy stellar mass - size \citep{Genel2018}, the color bimodality \citep{Nelson2018_2}, galaxy clustering \citep{Springel2018} and stellar mass function \citep{2018MNRAS.473.4077P}.

%These models are shown to reproduce several of the $z=0$ basic galaxy scaling relations, including the stellar mass - size \citep{Genel2018}, the color bimodality \citep{Nelson2018} and galaxy clustering \citep{Springel2018}, among others. As such, they are representative of the present day population of galaxies in the universe and reproduce the main environmental trends observed in satellites. Of particular relevance to this work, the abundance of low-mass and intermediate mass galaxies seems consistent with observationally estimated stellar mass functions \citep[e.g.][]{Pillepich2018MNRAS,Vazquez-Mata2020, Engler2021}.

%%%%%%%%%%%%%%%%%%%%%%%%%%%%%%%%%%%%%%%%%%%%%%%%%%%%%%%%%%
\begin{figure*}
	\includegraphics[width=2.05\columnwidth]{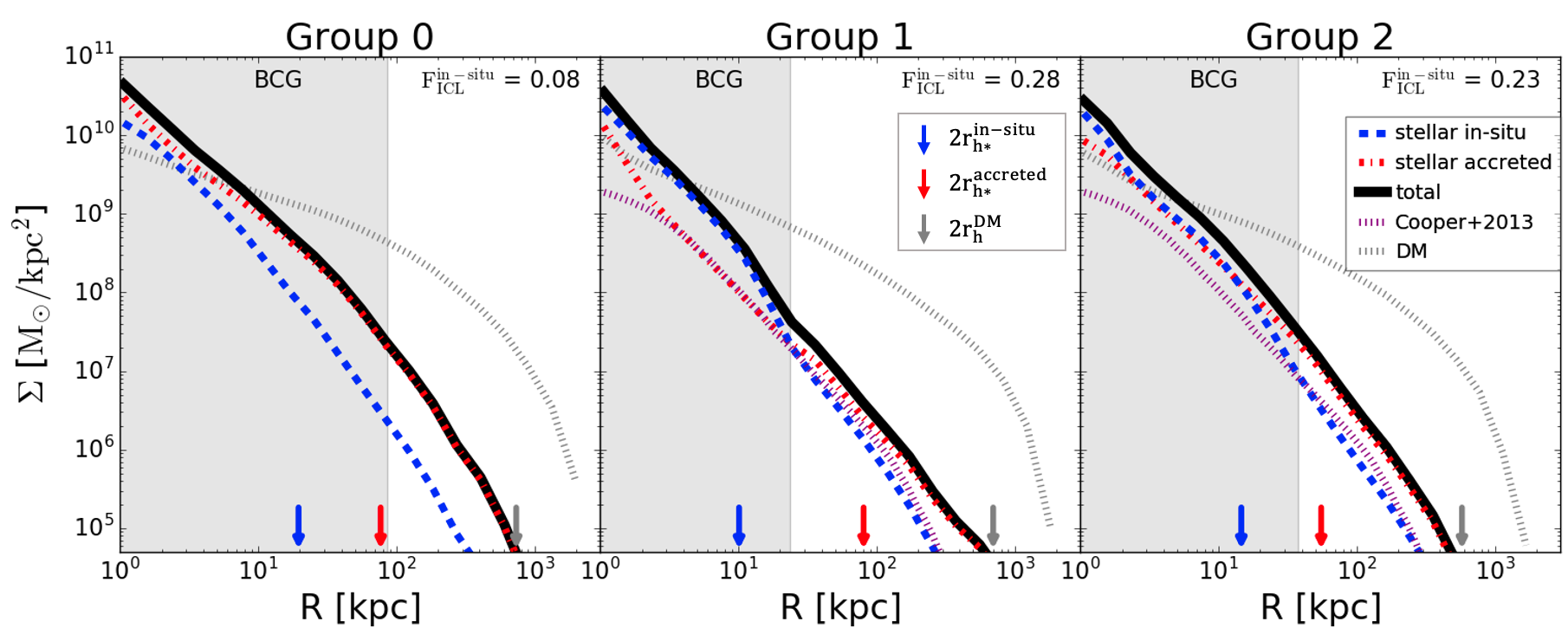}
    \caption{Projected density profiles of distinct components within the three most massive groups in TNG50 (including the BCG and ICL region and excluding satellites). The solid black curves represent the total stellar component, while the red curves depict the ex-situ (merged) stellar component, and the blue curves illustrate the in-situ stellar component. Dotted black lines trace the dark matter component, and purple dotted lines represent the median stellar mass surface density profiles from \protect \cite{Cooper2013}, based on a simplified particle tagging method. These profiles are shown only in the middle and right panels, where the halo masses match their sample. The shaded area delineates the region associated with the BCG, defined within $2 \, r_{h_{*}}$ for the central of each group. Arrows in matching colors indicate twice the half-mass radius of each component.}
    \label{fig1:density_profile}
\end{figure*}
%%%%%%%%%%%%%%%%%%%%%%%%%%%%%%%%%%%%%%%%%%%%%%%%%%%%%%%%%%

Galaxies in the simulations are identified using {\verb'SUBFIND'} \citep{2001MNRAS.328..726S, 2009MNRAS.399..497D} on Friends-of-Friends (FoF) groups \citep{1985ApJ...292..371D} identified with a linking length set to 0.2 times the mean interparticle separation. The object at the center of the gravitational potential of each group is called the ``central'' galaxy, while all other substructures are ``satellites''. The time evolution of galaxies and halos through the 99 snapshots of the simulation is followed using the SubLink merger tree algorithm \citep{2015MNRAS.449...49R}.

%%%%%%%%%%%%%%%%%%%%%%%%%%%%%%%%%%%%%%%%%%%%%%%%%%%%%%%%%%
\begin{table}
 \caption{Quantities related to the 3 most massive groups.}
 \label{tab:table1}
 \begin{tabular}{lccc}
  \hline
   & Group 0 & Group 1 & Group 2\\
  \hline
  M$_{200}$ \hspace{0.675cm}[$\mathrm{M}_{\odot}$] & $1.8 \times 10^{14}$ & $9.4 \times 10^{13}$ & $6.5 \times 10^{13}$\\
  R$_{200}$ \hspace{0.725cm}[kpc] & $1198.91$ & $959.61$ & $846.68$\\
  M$_\mathrm{BCG}^{*}$ \hspace{0.5cm}[$\mathrm{M}_{\odot}$] & $3.6 \times 10^{12}$ & $9.7 \times 10^{11}$ & $1.1 \times 10^{12}$\\
  r$_\mathrm{h_{*}}$ \hspace{0.925cm}[kpc] & $43.25$ & $11.75$ & $18.79$\\
  M$_\mathrm{ICL}^{*}$ \hspace{0.625cm}[$\mathrm{M}_{\odot}$] & $1.8 \times 10^{12}$ & $5.8 \times 10^{11}$ & $5.9 \times 10^{11}$\\
  M$_\mathrm{ICL}^{*,\mathrm{in-situ}}$ [$\mathrm{M}_{\odot}$] & $1.4 \times 10^{11}$ & $1.6 \times 10^{11}$ & $1.3 \times 10^{11}$\\
  f$_\mathrm{ICL}$ & $0.48$ & $0.60$ & $0.56$\\
  F$_\mathrm{ICL}^\mathrm{in-situ}$ & $0.08$ & $0.28$ & $0.23$\\
  \hline
 \end{tabular}
\end{table}
%%%%%%%%%%%%%%%%%%%%%%%%%%%%%%%%%%%%%%%%%%%%%%%%%%%%%%%%%%

\subsection{Defining ICL and cleaning method}
\label{ssec:cleaning}

We focus the analysis on the $3$ most massive systems included in the TNG50 box, which fulfill  $M_{200}/\rm M_{\odot} > 5 \times 10^{13}$, comparable to nearby systems such as the Fornax and the Virgo clusters.  Virial masses for these systems are listed in table~\ref{tab:table1}, where virial quantities are defined at the virial radius $\rm r_{200}$, corresponding to the distance at which the enclosed density is $200$ times the critical value.

To define the particles that are considered part of the ICL we proceed as follows. For each of our groups, we remove all stars that {\verb'SUBFIND'} considers bound to surviving satellite galaxies. The left panel of Fig.~\ref{fig:sat_stellar} shows an illustration of this component for our Group 1. Furthermore, to avoid possible contamination by stars in the outskirts of the surviving satellites that are wrongly assigned to the host halo by {\verb'SUBFIND'}, we further apply a ``cleaning'' of our ICL: we remove any stellar particle within $4 \, \rm r_{h_{*}}$ of the center of any subhalo with $M_* > 10^8$\msun, where $\rm r_{h_{*}}$ refers to the stellar half mass radius of the subhalo. This ensures that any stars from satellites and their surrounding stellar halo are not considered part of the ICL.

This leaves us with a sample consisting of all the remaining stellar particles in the FoF group that are beyond $2 \, \rm r_{h_{*}}$ of the central galaxy and within the virial radius $r_{200}$ of the host cluster, which we call ICL.  
Stars in the ICL are further divided into those born ``in-situ'' and those ``accreted''. Following \citet{Rodriguez2016}, our algorithm checks if the star is bound to the main progenitor of the central galaxy at the time of birth (or the closest snapshot to it), in which case it is labeled as in-situ, or it is otherwise associated by {\verb'SUBFIND'} to a different substructure in the SubLink merger tree, in which case it is labeled as accreted. The catalogs for in-situ/accreted are provided within the IllustrisTNG database. As an illustration, the middle and right panels of Fig.~\ref{fig:sat_stellar} show the accreted and in-situ components of the ICL in Group 1, respectively.

To quantify the mass within the ICL and, specifically, the in-situ stellar component, we introduce the fractions $f_\mathrm{ICL}$ and $F_\mathrm{ICL}^\mathrm{in-situ}$. Various definitions of these fractions exist in the literature, and for the purpose of our study, we adopt the following definitions:

\begin{equation}
 f_\mathrm{ICL}=\frac{M_\mathrm{ICL}^{*}}{M_\mathrm{BCG}^{*}},
\end{equation}
and
\begin{equation}
 F_\mathrm{ICL}^\mathrm{in-situ}=\frac{M_\mathrm{ICL}^{*,\mathrm{in-situ}}}{M_\mathrm{ICL}^{*}},
\end{equation}
where $M_\mathrm{ICL}^{*}$ refers to the mass of the stellar component in the ICL, $M_\mathrm{ICL}^{*,\mathrm{in-situ}}$ represents the mass of the in-situ stellar component in the ICL, and $M_\mathrm{BCG}^{*}$ is the stellar mass within $2 \, \rm r_{h_*}$ of the central galaxy. The values for these quantities in the three most massive systems in TNG50 are presented in table~\ref{tab:table1}.

\subsection{H$_{\alpha}$ predictions}\label{HPredictions}

%%%%%%%%%%%%%%%%%%%%%%%%%%%%%%%%%%%%%%%%%%%%%%%%%%%%%%%%%%
\begin{figure*}
	\includegraphics[width=2.05\columnwidth]{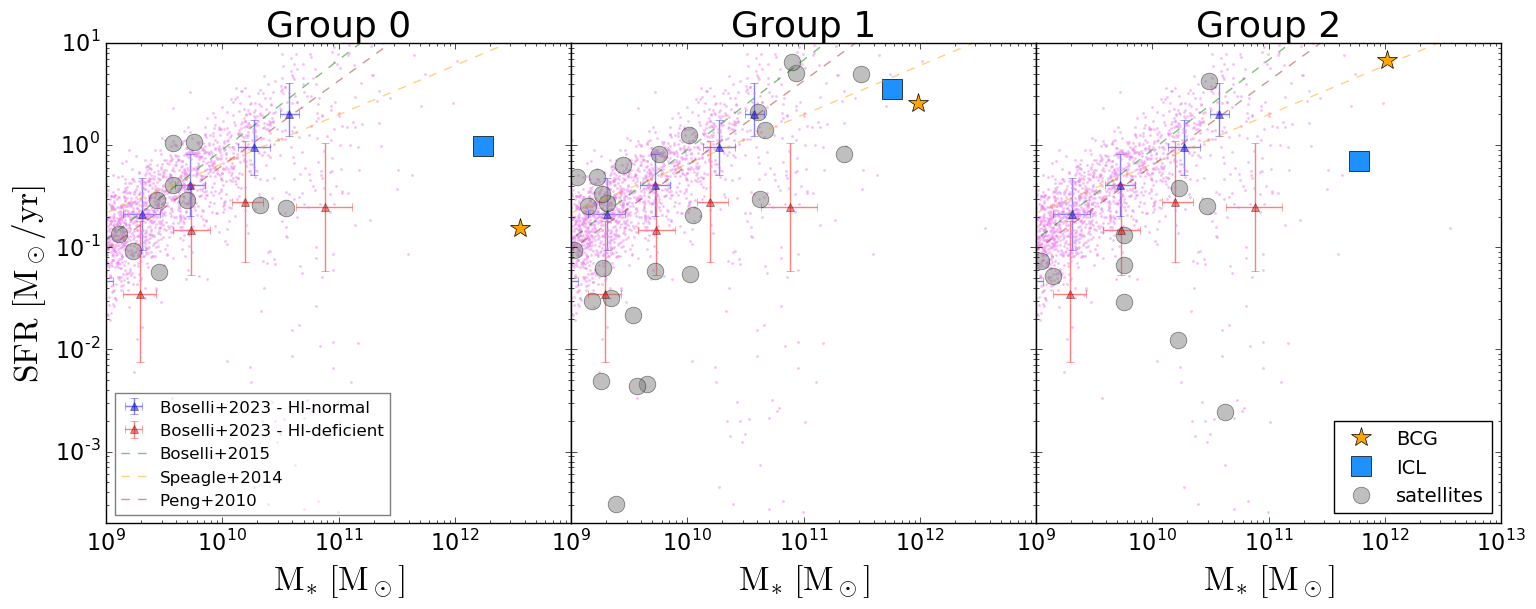}
    \caption{Star formation rate (SFR) plotted against stellar mass for various components within each group. Results for the BCGs are represented by orange stars, those for the ICL component by blue squares, and those for satellites within the ICL region by grey circles. Violet points illustrate the SFR distribution for central galaxies in TNG50 within the mass range limit. For comparison the main sequence relation for the galaxies in Virgo cluster from \cite{Boselli2023} are shown by blue and red markers, the extrapolated relations for late-type nearby galaxies are shown by green dashed line \citep{Boselli2015}, the relation for a large sample of SDSS galaxies are shown by brown dashed line \citep{Peng2010}, and work by \cite{Speagle2014} is presented in orange dashed line.}
    \label{fig4:SFR_Mstr}
\end{figure*}
%%%%%%%%%%%%%%%%%%%%%%%%%%%%%%%%%%%%%%%%%%%%%%%%%%%%%%%%%%

In order to compare results with the recent observations of the ICL in nearby clusters (i.e. Virgo), we use the calibration of \cite{2012} to convert the star formation rates to H$_{\alpha}$ luminosities. These calibrations are based on evolutionary population synthesis models, in which the emergent spectral energy distributions (SEDs) are derived for synthetic stellar populations with a prescribed age mix, chemical composition, and \cite{Kroupa2001} initial mass function (IMF). The resulting relation is
\begin{equation}
 \log L_\mathrm{x}  = \log (\dot{M}_{*} / \mathrm{M}_{\odot} \, \mathrm{yr}^{-1}) + \log C_\mathrm{x},
 \label{eqHalpha}
\end{equation}
where $L_\mathrm{x}$ denotes the $H_{\alpha}$ luminosity in units of $\mathrm{erg} \,\, \mathrm{s}^{-1}$, $\dot{M}_{*}$ represents the star formation rate in units of $\mathrm{M}_{\odot} \, \mathrm{yr}^{-1}$, and $C_\mathrm{x}$ is the logarithmic SFR calibration constant which we set to be 41.27 \citep{2012,2011, 2011ApJ...741..124H}.

\section{Results}\label{result}

We begin by characterizing the mass distribution within our groups, focusing on the projected radial profiles to facilitate comparison with observational studies. Fig.~\ref{fig1:density_profile} illustrates the projected radial distribution of stars (solid thick black line), and dark matter (black dotted line) after projecting each group along the $x$ axis. The stellar component is predicted to fall off more steeply than the dark matter component. For comparison, the purple dotted line represents the stellar mass surface density profiles from \cite{Cooper2013}, based on a simplified particle tagging method, averaged for halos with $10^{13.5} < M_{200}/\mathrm{M}_\odot < 10^{14}$. These results align closely with our stellar profiles (solid black lines) within the ICL region for the two less massive clusters (as shown in the right and middle panels). However, in the case of our most massive cluster, having a halo mass twice that of the most massive halo in their sample, differences with the predictions in \cite{Cooper2013} are expected (as a result the dashed purple line is removed from this panel). Note that the tagging method in \cite{Cooper2013} does not properly account for the in-situ star formation in the BCG, therefore giving a flatter slope in the inner regions. We further divide the stars according to their origin: in-situ (represented by the thin blue dot-dashed curve) and accreted (depicted by the thin red long-dashed curve). The relatively substantial contribution of the in-situ population to the ICL at hundreds of kpc is intriguing. In our systems, in-situ born stars contribute between $8\%$ and $28\%$ of the stellar mass in the ICL (refer to the top right corner of each panel for the specific percentage of each group). 

An inspection of the birth place of the in-situ labelled stars in the ICL shows that the majority of them are formed already at large distances from the BCG (although not associated with substructures) and are not stars from the central galaxy kicked-off by mergers, a common formation path for in-situ stars in the diffuse stellar halos of MW-like galaxies, and as is the case for our most massive cluster (which experienced a recent major merger in its formation history). However, for the remaining groups in our sample, over 60\% of the in-situ ICL is formed at $r > 40$ kpc or $r > 2 \, \rm r_{h_{*}}$ (where distance is in physical kpc and is measured with respect to the BCG or its main progenitor at the time of birth of the star). This suggests that gas in the ICM must somehow become available for star formation beyond the inner regions dominated by the BCG. This star formation mode is thus happening in unconventionally diffuse conditions, very different from the star formation resulting from the dense interstellar medium of galaxies.

We calculate the total SFR of ICM gas at $z=0$ for our groups and compare this to the level of star formation ongoing in the central BCG as well as in other galaxies. Fig.~\ref{fig4:SFR_Mstr} highlights that, taken as a unit, the ICL (integrated within the virial radius) is forming stars at rates comparable to that of the BCG and other massive galaxies. For this comparison, we assign a stellar mass to the ICL equal to the sum of all in-situ and accreted stellar particles that lie beyond $2 \, \rm r_{h_{*}}$ and up to the virial radius and that are not associated to any satellite or identified substructure by {\verb'SUBFIND'}. Similarly, we compute its SFR by summing the star formation rates of all gas cells that are not gravitationally bound to any substructure within the same region. Fig.~\ref{fig4:SFR_Mstr} shows that while the ICL lies systematically below the extrapolation of the ``main sequence'' defined by lower-mass star forming galaxies (magenta points show central galaxies within the TNG50 box), its SFR is comparable to many of the massive satellite galaxies in the group (gray circles) and above the SFR of the central BCG in the case of Group 0 and Group 1 (left and middle panels). For comparison, we present the extrapolated ``main-sequence'' relation derived from the galaxies in the Virgo cluster (see \citealt{Boselli2023}), for a sample with normal HI gas content (HI-normal) and deficient HI gas content (HI-deficient); the mean values and standard deviations for HI-normal and HI-deficient objects are shown by blue and red markers. We are also presenting results from late-type nearby galaxies in the Herschel Reference Survey by \citealt{Boselli2015} (dashed green line), a large statistical sample of SDSS galaxies from \citealt{Peng2010} (dashed brown line), and the extrapolated time-dependent main-sequence relation derived by \citealt{Speagle2014} (dashed orange line). One should also keep in mind that the fraction of quenched galaxies produced by the TNG model might overestimate the star formation in massive galaxies \citep[e.g., see Fig. 8 in][]{Donnari2021b}.

\begin{figure}
	\includegraphics[width=\columnwidth]{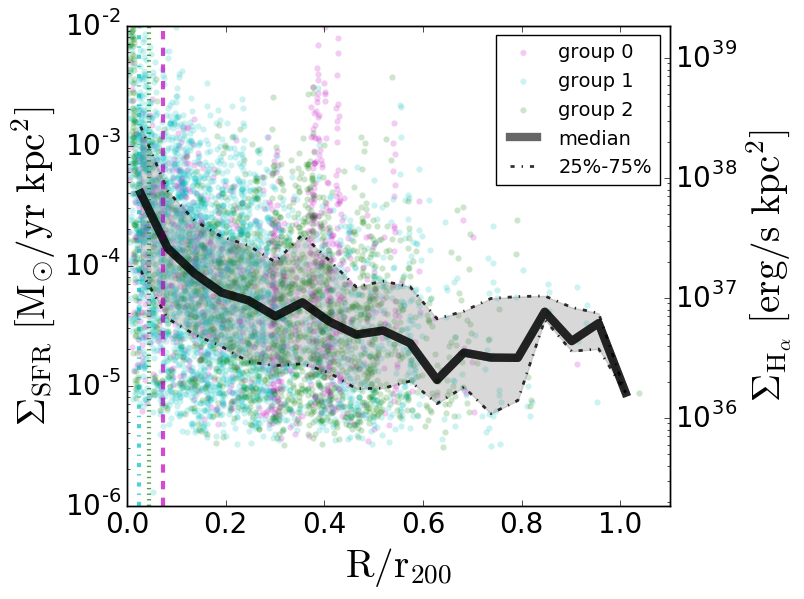}
    \caption{Star formation rate surface density profiles for the three most massive groups in TNG50. Different colored points represent different groups, with individual points of the same color illustrating the scatter between various regions at a given clustercentric distance for each group (star formation is measured within square bins of $2 \times 2$ [kpc $\times$ kpc] when projecting each group along the $x$-axis). Vertical lines, color-matched to each group, indicate $2 \, \rm r_{h_{*}}/ \rm r_{200}$. The black solid line represents the median trend, and the grey shaded region corresponds to the 25-75 percentile dispersion of $\Sigma_{\rm SFR}$ as a function of clustercentric distance for all groups (the median and dispersion are only calculated from regions that have non-zero star formation).}
    \label{fig3:SFR_profile}
\end{figure}

The ongoing star formation in the ICM shows a decreasing trend with distance from the center of the cluster. Fig.~\ref{fig3:SFR_profile} presents the surface density of star formation rate ($\Sigma_{\rm SFR}$), computed by projecting our groups along the $x$-axis and integrating star formation within square bins of $2$ kpc on each side. Different colors represent different groups, while individual points of the same color depict the scatter between various regions within a given group. Bins with no star formation at all are not shown and not considered for calculating the median and the corresponding dipersion. The thick solid curve corresponds to the median trend (considering all groups collectively) of $\Sigma_{\rm SFR}$ as a function of clustercentric distance, highlighting that the inner regions of the ICM exhibit more prominent star formation compared to the outskirts. We have checked that the vertical magenta structures in group 0 correspond to a few dominant clumps in the ICL with a wide range in predicted SFR.

%%%%%%%%%%%%%%%%%%%%%%%%%%%%%%%%%%%%%%%%%%%%%%%%%%%%%%%%%%
%%%%%%%%%%%%%%%%%%%%%%%%%%%%%%%%%%%%%%%%%%%%%%%%%%%%%%%%%%
%%%%%%%%%%%%%%%%%%%%%%%%%%%%%%%%%%%%%%%%%%%%%%%%%%%%%%%%%%
\begin{figure*}
	\includegraphics[width=2.05\columnwidth]{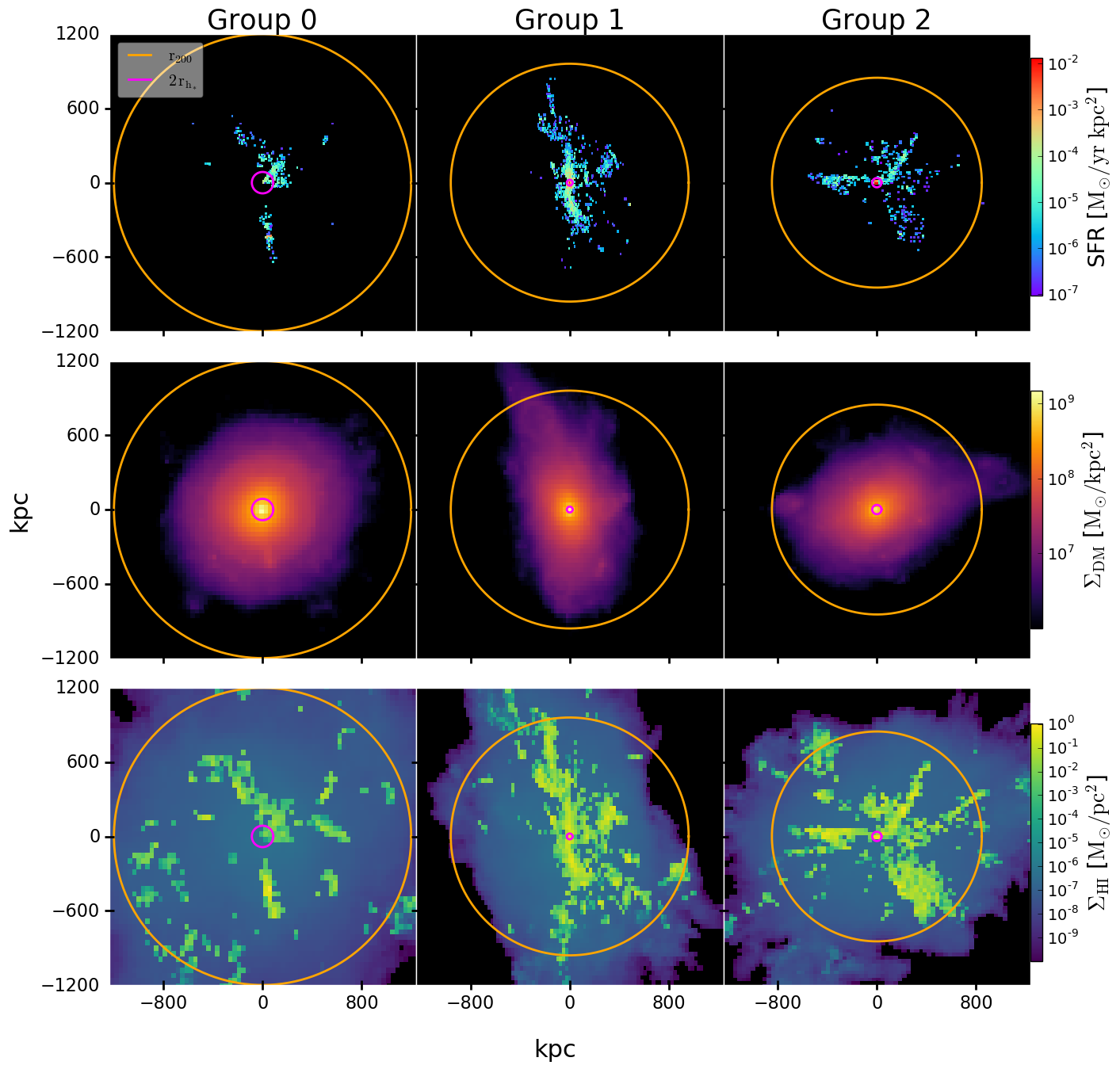}
    \caption{Projected maps of star-forming gas, the dark matter component and HI gas in the three most massive groups in TNG50, presented in columns from left to right for groups 0, 1 and 2. Rows from top to bottom show: SFR, dark matter surface density, and HI gas surface density. The colorbar on the right side of each row indicates the respective quantities displayed. The ICL is highlighted by bounds marked with magenta and orange circles, representing $2 \, \rm r_{h_{*}} <r< \rm r_{200}$. These results underscore the widespread nature of star formation, organized in filamentary structures within the halo of our targeted groups. }
    \label{fig:xyz}
\end{figure*}
%%%%%%%%%%%%%%%%%%%%%%%%%%%%%%%%%%%%%%%%%%%%%%%%%%%%%%%%%%
To ease comparison with observations, we convert $\Sigma_{\rm SFR}$ into a surface density of H$_\alpha$ luminosity (right vertical axis) as explained in detail in Sec.~\ref{HPredictions} and following \cite{2012}. Typical values in the ICM of our sample span $\Sigma_{\mathrm{H}_\alpha} \sim 10^{38} \rm - 10^{36}$\unitHa\; with considerable scatter. These values are lower than the average measured $\Sigma_{\mathrm{H}_\alpha} \sim 10^{40}$\unitHa\; in central regions of disk galaxies, but comparable to the values in the outskirts of MW- and LMC-like objects \citep{Tacchella2022, Bundy2015}. Beyond clustercentric distances $R$ $> 0.8 \, \rm r_{200}$ diffuse star formation is rare in all our systems.

Projected maps of our halos in Fig.~\ref{fig:xyz} show in more detail where the star formation in the intracluster space is happening. Panels on the first row show the star-forming gas in the region $2 \, \rm r_{h_{*}} <r< \rm r_{200}$ (bounds shown with magenta and orange circles) color coded by SFR (color map on the right). %\lvs{The bin size is 3x3 kpc$^2$ and values of SFR reported correspond to the cumulative SFR of all star-forming gas cells projected in the bin. @Niusha: can you check that the bin size is correct?}. 
Star formation at $z=0$ is widespread in these systems and follows filament-like patterns. We highlight that special care was taken to remove any material gravitationally associated with surviving satellites, as described in Sec.~\ref{ssec:cleaning}, and this star formation is occurring in gas seemingly associated to the central host halo and not to bound substructures. In fact, further inspection of 3D projections for each system revealed that the filamentary nature of the star-forming gas follows the distribution of neutral gas in the main host halo (bottom row), and looks loosely linked to the dark matter distribution (middle row) for at least 2 of the systems (groups 1 and 2). 

%%%%%%%%%%%%%%%%%%%%%%%%%%%%%%%%%%%%%%%%%%%%%%%%%%%%%%%%%%
\begin{figure*}
	\includegraphics[width=2.05\columnwidth]{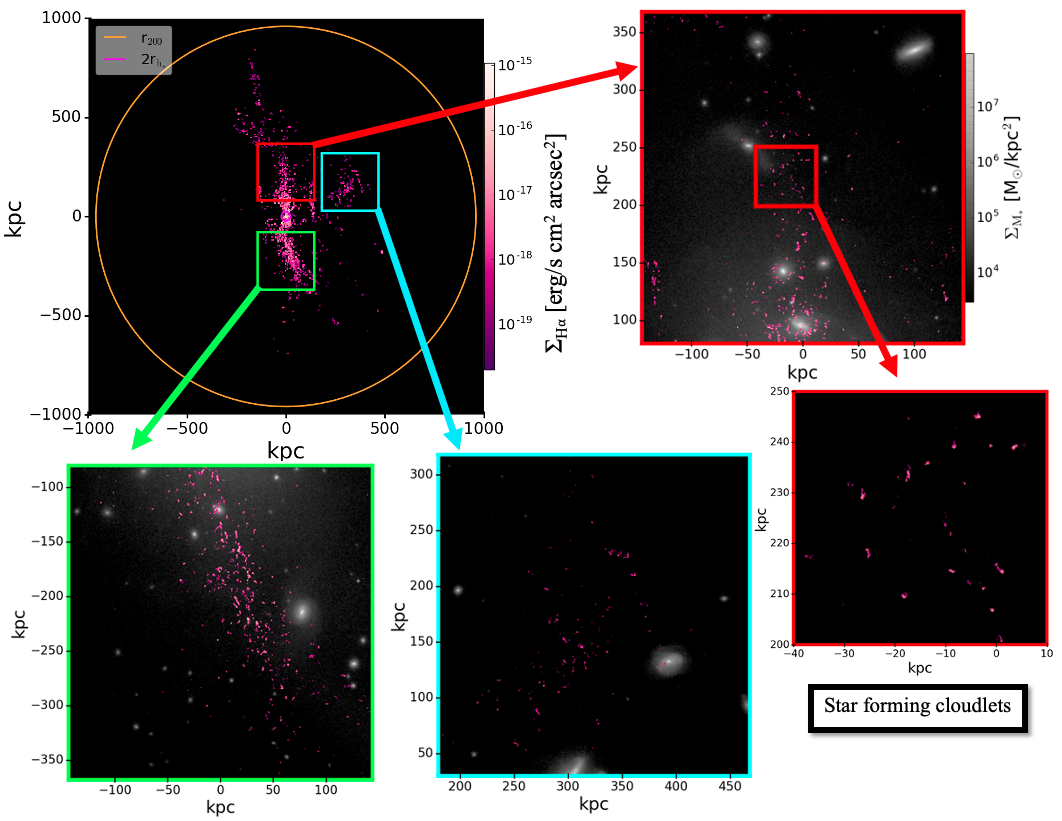}
    \caption{Main Panel: $x$--$y$ projection of star-forming gas in Group 1, color-coded by the flux of H$_\alpha$ and smoothed to 1 arcmin, revealing the filamentary structure of the star-forming gas. Magenta and orange circles represent regions at $2 \, \rm r_{h_{*}}$ and $\rm r_{200}$. Zoomed-in Panels: H$_\alpha$ flux of star-forming gas (smoothed to 3 arcsec), with gray color map representing stellar distribution. These findings suggest that star-forming regions are not necessarily linked to ram-pressure tails from gas-rich galaxies. Instead, stars appear to be forming within cloudlet-like structures, condensing along filamentary structures within the ICM.}
    \label{fig:zoom}
\end{figure*}
%%%%%%%%%%%%%%%%%%%%%%%%%%%%%%%%%%%%%%%%%%%%%%%%%%%%%%%%%%

To gain further insight, we zoom into smaller regions of group 1 in Fig.~\ref{fig:zoom}. The gray color map shows the distribution of the stellar component in each box and reveals the presence of several galaxies, some with associated stellar tails---most prominently observable in the two innermost regions (red and green panels). We highlight the presence of star forming gas colored in magenta by their estimated $\mathrm{H}_\alpha$ flux (refer to the color bar on the side of the main panel). For the zoomed-in panels, the color maps are calculated in $\sim 0.24 \times 0.24$~kpc bins to align with the $3$ arcsec spatial resolution of the VESTIGE survey of the Virgo cluster \citep{Boselli2018}, while the main panel is smoothed to 1 arcmin.

The zoomed-in regions in Fig.~\ref{fig:zoom} show that the majority of the star forming gas is mostly unrelated to the distribution of stars or galaxies. Star forming regions are also not obviously associated to ram-pressure tails from gas-rich galaxies. Note that this is different from what has been observed in cases such as \cite{Smith2010, Mihos_2016, Boselli2018, Gullieuszik2020}, where the association of the star-forming region to tails of ram-pressure stripped galaxies is clear upon visual inspection. While these ram-pressure tails have been found in simulations including TNG50 \citep{Rohr2023}, they are not the main contributor to the star forming gas in our systems. Instead, stars seem to be forming in cloudlet-like objects with kpc typical size, which condensate along filamentary structures in the ICM. 

%%%%%%%%%%%%%%%%%%%%%%%%%%%%%%%%%%%%%%%%%%%%%%%%%%%%%%%%%%
\begin{figure*}
	\includegraphics[width=2.05\columnwidth]{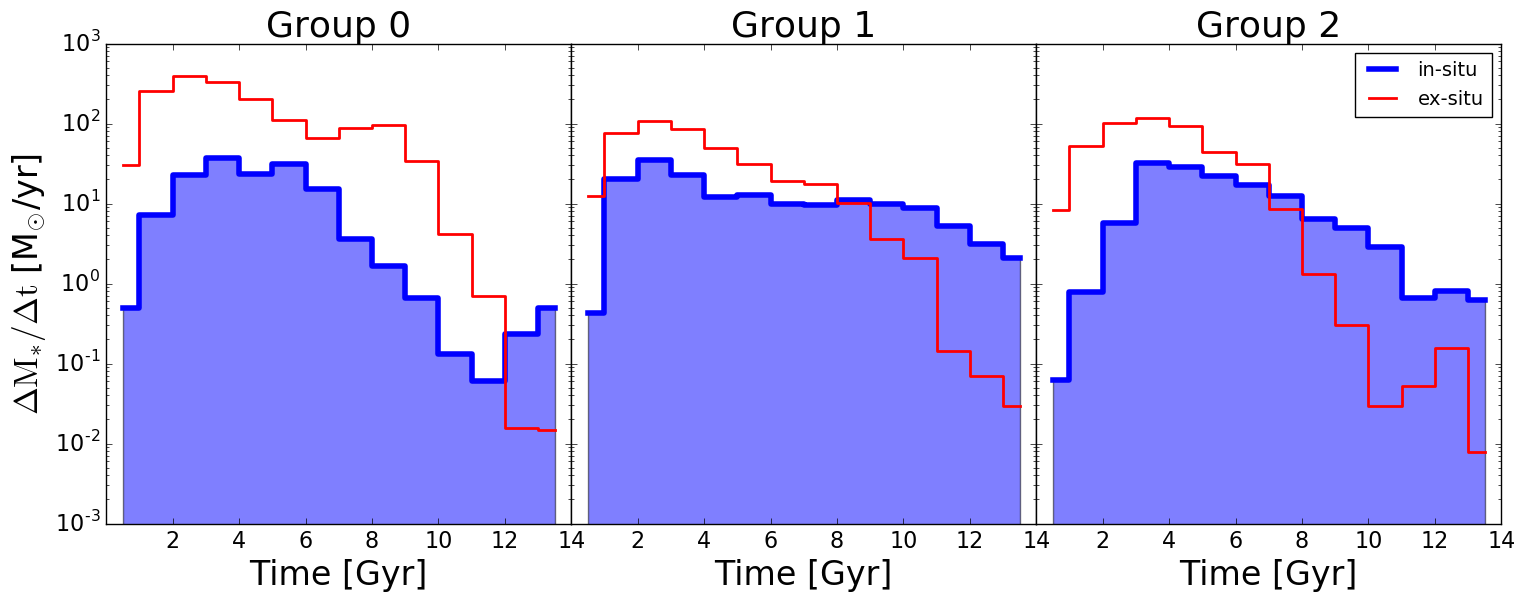}
    \caption{Average star formation rate for the in-situ (blue shaded histogram) and merged (red histogram) components of the ICL. The in-situ intra-cluster star formation is a continuous phenomenon over time, with an average peak intensity around $z \sim 2 - 3$ (cosmic times $2-4$ Gyr).} 
    \label{fig5:age_hist}
\end{figure*}
%%%%%%%%%%%%%%%%%%%%%%%%%%%%%%%%%%%%%%%%%%%%%%%%%%%%%%%%%%

This is not the first time that small-scale cold clouds in the ICM have been reported in TNG50. In a detailed analysis of the ICM in $z=0.5$ halos, \citet{Nelson2020} highlighted this fundamental prediction from the TNG model in combination with the high resolution achieved by TNG50. These small-scale cold clouds seem consistent with being seeded by local inhomogeneities in gas density that later grow due to increased cooling. While the role of magnetic fields in forming these clouds is not clear, the magnetic pressure in these clouds is large and exceeds the thermal pressure, providing support and making these features long-lived, on the scale of several Gyrs. Interestingly, the results in this work show that some of these cold clouds manage to become star-forming and contribute a non-negligible fraction of the stars to the diffuse ICL component.  

In agreement with the findings of \citet{Nelson2020}, the metallicity of in-situ born ICL stars (with $r_{\rm born} > 2 \, \rm r_{h_{*}}$) is relatively high, registering a median value of $[Fe/H] \sim -0.77$ with a substantial scatter of approximately $0.39$ dex, comparable to the accreted ICL, the metallicity of dwarf galaxies with $M_* \sim 10^8$\msun, and the outskirts of more luminous, MW-like galaxies. This suggests that the gas seeding these density perturbations has galaxy-like abundances and it likely comprises gas that has, a long time ago, been stripped from infalling galaxies. Obvious associations to those satellites do not seem to persist until $z=0$ in our simulations, suggesting that the timescales for the clouds to form and become star-forming are comparable to the orbital times of the satellites. The similarity between the metallicities for the in-situ and accreted components complicates the chances of observationally distinguishing these two populations.

\begin{figure*}
	\includegraphics[width=2.05\columnwidth]{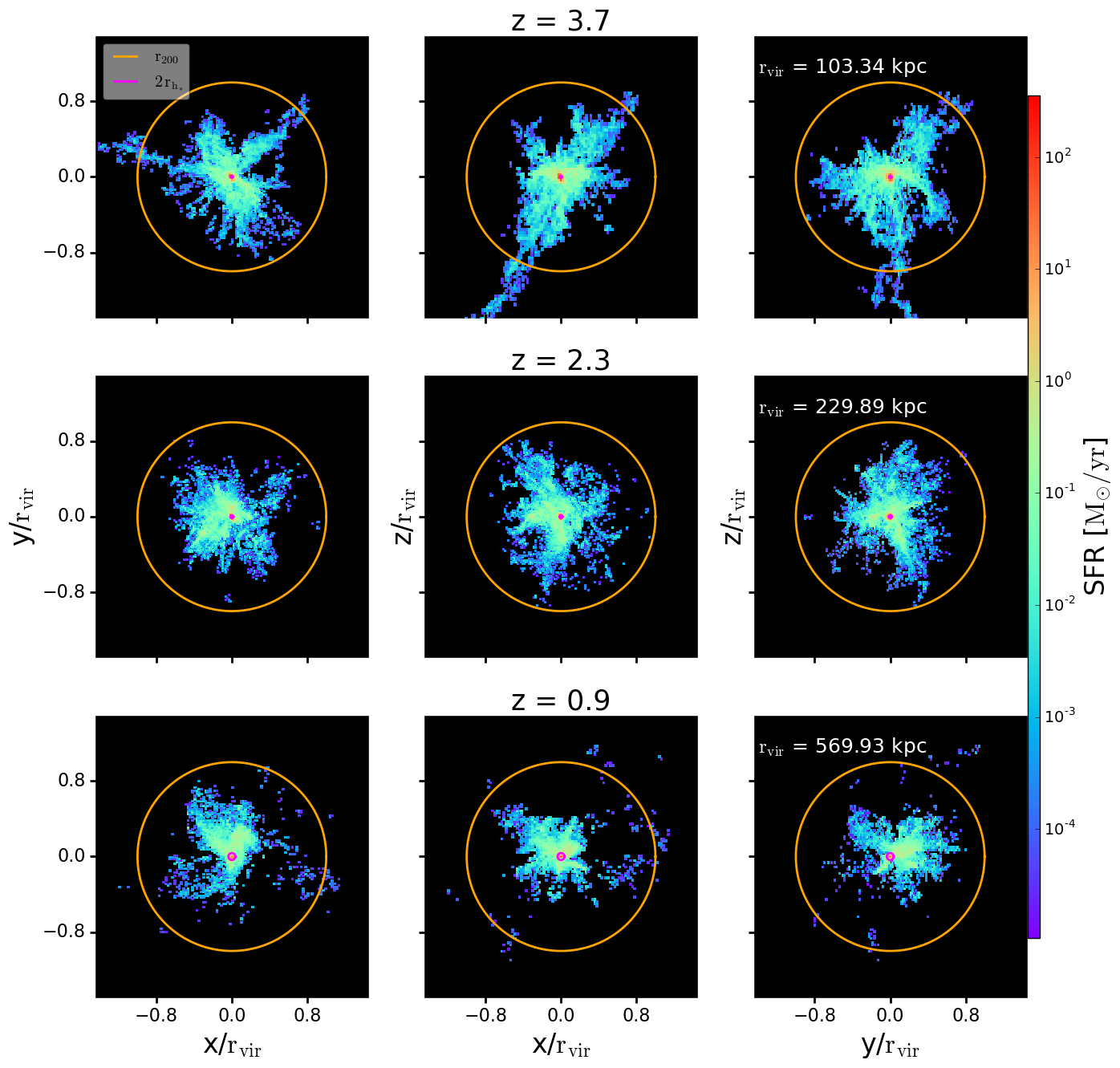}
    \caption{Projected maps of star-forming gas at different redshifts for Group 0, arranged in columns from left to right: x-y, x-z, and y-z projections. The projections are normalized by the virial radius of the cluster at each redshift, with the specific value provided in the top left corner of the last panel in each row. Rows from top to bottom correspond to redshifts $z = 3.7,\, 2.3,\, 0.9$. The colorbar on the right side indicates the star formation rate (SFR) in each region. The ICL is highlighted by bounds represented by magenta and orange circles, signifying $2 \, r_{h_{*}} < r < \rm r_{200}$. These results emphasize the presence of a diffuse mode of star formation within the ICL of our targeted groups, even at earlier redshifts.}
    \label{fig:SFR_highz}
\end{figure*}

The physical conditions for the star-forming clouds seem comparable to those of normal star-forming regions in the ISM of galaxies. The median density of the star-forming cells flagged as part of the ICM is $\sim 0.2 \; \rm cm^{-3}$, which is just above the density threshold for star formation in the TNG baryonic treatment $n \approx 0.1\; \rm cm^{-3}$ (Please see Appendix~\ref{SF_condition} for a more detailed comparison of density and temperature for star-forming cells in the ICL compared to those in satellite galaxies.) Interestingly, the fraction of stellar mass born in-situ in the ICM seems only weakly dependent on numerical resolution, as some of the clusters selected in TNG100 show similar levels of star formation rate in the ICM (see Appendix \ref{res_effect}). The presence of these in-situ born stars in the ICL is therefore a fundamental consequence (or prediction) of the model in several resolution levels.\\
%Interestingly, the fraction of stellar mass born in-situ in the \replaced{ICM}{ ICL} seems quite independent of numerical resolution, as clusters selected in TNG100 show similar levels of star formation rate in the \replaced{ICM}{ ICL} (see Appendix \ref{res_effect}). The presence of these in-situ born stars in the ICL is therefore not driven by the increase in resolution from TNG50 to TNG100 but is instead a more fundamental consequence (or prediction) of the model.  
At z=0, the velocity distributions of both the in-situ and accreted stellar components are found to be remarkably similar, suggesting significant virialization within the cluster. This hints at the possibility that the in-situ component must also be relatively old, comparable in age to the accreted stars. We also note that at the current integrated star formation rates shown in the intra-cluster region of our simulated clusters, SFR $\sim$ 1 $\rm \mathrm{M}_\odot$/yr (see Fig.~\ref{fig4:SFR_Mstr}), it would take $\sim 100$ Gyr to form the total mass of the in-situ ICL stars listed in Table~\ref{tab:table1}. This diffuse mode of star formation must therefore have also been present at earlier times, and with higher intensity. We confirm this by showing the distribution of stellar mass formed in-situ in a given time-bin in Fig.~\ref{fig5:age_hist}. This is calculated by adding all the stellar mass of the in situ-labeled stars in the ICL with a given age divided by the size of the time-bin to provide an average star formation rate. The blue shaded histograms indicate that the in-situ intra-cluster star-formation is a rather continuous phenomenon over time and was, on average, stronger around $z \sim 2$-$3$ (i.e., cosmic times $2$-$4$ Gyr). As an illustration, we show in Fig.~\ref{fig:SFR_highz} the predicted SFR maps for our highest mass group at three different redshifts: $z = 3.7, \, 2.3 \, \rm and \, 0.9$. In all cases, extended star forming areas appear out to almost the virial radius of the cluster progenitor. An examination of the birthplaces of the in-situ stars in the ICL reveals that a fraction of these stars may have originated in the central galaxy and were subsequently ejected due to merger events. This phenomenon is especially noticeable in one of our groups, which experienced a recent major merger in its formation history. However, for the rest of our sample, we estimate that the majority (over 60\%) of the in-situ ICL stars formed at $r > 40$ kpc or $r > 2 \, \rm r_{h_{*}}$, relative to the BCG or its main progenitor at the time of their formation.

These stars are forming from the cool gas present in the ICM of the clusters, as shown in the star-forming gas distributions at higher redshifts in Fig.~\ref{fig:SFR_highz}. The existence of cool gas in filamentary structures around these clusters at higher redshift ($z \gtrsim 2$) has also been recently reported in the study by \cite{Rohr2024}, which examines more massive clusters (with masses ranging from $10^{14.3}$ to $10^{15.4} \, \rm M_{\odot}$) from the TNG-Cluster simulation suite \citep{Nelson2024}. This suite employs the same baryonic physics as TNG50 but at a lower resolution. Our measured SFRs in the ICM for the mass range of our three systems are roughly in agreement with their reported SFRs at $z\sim 0.5$ where the mass range of their more massive clusters overlaps with our systems.

%%%%%%%%%%%%%%%%%%%%%%%%%%%%%%%%%%%%%%%%%%%%%%%%%%%%%%%%%%
%%%%%%%%%%%%%%%%%%%%%%%%%%%%%%%%%%%%%%%%%%%%%%%%%%%%%%%%%%

%\newpage
Have these star-forming regions been detected in observations?  In the nearby universe, ongoing star formation in the tails of ram-pressured stripped galaxies has been observed in the past, for example in the Virgo cluster \citep{Mihos_2016, Boselli2018}. However, the predictions from TNG50 are different in nature from these observations, since the star-forming cloudlets are currently not associated with any satellite. In that sense, they are closer to a recently identified class of young, isolated, and star-forming clouds in Virgo, see \cite{Jones2024, Beccari2017}. We cannot rule out that this gas originated as ram-pressure tails stripped from galaxies in the distant past, where any possible correlation with the galaxy distribution has been long erased. However, if that is the case, the timescale for turning this gas into stars seems too long compared to observations of ram-pressure tails which can still be traced back to their original galaxy. Therefore we consider this a different process than what is observed in ram-pressure tails.

The color coding of the gas in Fig.~\ref{fig:zoom} indicates that, at $z=0$, the expected H$_\alpha$ surface brightness for the intra-cluster gas is rather modest, $\Sigma_{\rm H_\alpha} \sim 1.6 \times 10^{-19}$--$2.6 \times 10^{-18}$ $\rm erg\;\rm s^{-1}\; cm^{-2}\; \rm arcsec^{-2}$, which could partially explain why this phenomena has so far escaped detection\textsuperscript{2}\footnote{These quoted values for $\rm H_\alpha$ brightness are derived from Fig.~\ref{fig:zoom}, where we binned the star-forming gas at similar spatial resolution as VESTIGE (3 arcsec) and measured the $\rm H_\alpha$ flux within each 'pixel``. If we analyze the results and compute the median and dispersion, we obtain a median value of $6.8 \times 10^{-19}$$\rm erg\;\rm s^{-1}\; cm^{-2}\; \rm arcsec^{-2}$ and a dispersion ranging from $1.6 \times 10^{-19}$--$2.6 \times 10^{-18}$ $\rm erg\;\rm s^{-1}\; cm^{-2}\; \rm arcsec^{-2}$.}. For comparison, the most extended and deepest surveys available, such as VESTIGE, have a sensitivity $\Sigma_{\rm H_\alpha} \sim 2 \times 10^{-18}$ $\rm erg\;\rm s^{-1}\; cm^{-2}\; \rm arcsec^{-2}$ at 3 arcsec angular resolution \citep{Boselli2018}\textsuperscript{3}\footnote{The stacking of pixels to increase image sensitivity of the images should be limited to reasonable scales. The 3 arcsec scale (~0.24 kpc at the distance of the Virgo cluster) mentioned in the VESTIGE paper is optimized for detecting structures associated with the tails of stripped gas in galaxies. While stacking pixels on 100 kpc scales could theoretically increase sensitivity as $\sqrt(N)$ (where N is the number of pixels), this assumes a perfectly flat background, which is rarely the case. Issues such as large-scale structures from flat fielding imperfections and correlated noise due to bright stars and unwanted gradients from non-uniform illumination can affect the results (see Fig. 4 of \cite{Boselli2018}, VESTIGE paper I). Moreover, narrow-band filters centered on the rest-frame $\rm H_\alpha$ line are also sensitive to $\rm H_\alpha$ emission from Galactic cirrus. Hence the contribution of the Galactic cirrus in the direction of the Virgo cluster can complicate measurements (see Fig. 16 of the same VESTIGE I paper). While IFU spectroscopy could help confirm these features, technical limitations exist. For instance, MUSE/VLT can achieve  $\Sigma_{\rm H_\alpha} \sim  10^{-18}$ $\rm erg\;\rm s^{-1}\; cm^{-2}\; \rm arcsec^{-2}$ on $\sim$1 arcsec scales in $\sim$1h integration, but its 1 arcmin FoV is insufficient for mapping large areas like the Virgo cluster ($\sim$100 sq.deg), necessitating pre-selection of observation targets.}. On the other hand, there are claims of detection of runaway cooling of the hot gas in high-$z$ clusters leading to star formation events with SFR $\sim$ 900 $\rm \mathrm{M}_\odot$/yr offset from the main host galaxy by $\sim 25$ kpc for the coolest gas and $\sim 50$ kpc for the X-ray peak \citep{Hlavacek2020} that might be comparable to some of the in-situ formed stars in our simulations. 

Finally, we must also consider the possibility that these in-situ formed ICL stars are a numerical effect born out of specific choices made for the baryonic treatment in the simulation. While the multi-phase nature of the gas in the outskirts of groups and clusters has been reported before by several simulation groups \citep{Nelson2018,Butsky2019, Oppenheimer2021}, whether warm/cold gas cells in the ICM can form stars is less robustly predicted by numerical codes. The fact that most star-forming clouds sit near the density threshold for star formation suggests that increasing the value of that parameter might significantly lower the occurrence of stars born in the in-situ ICL component, but more rigorous tests must be performed to quantify this effect. More detailed baryonic prescriptions that explicitly resolve the multiphase structure of the gas in the ISM of galaxies advocate for larger density thresholds for star formation, $n \sim 1$ - $100$ $\rm cm^{-3}$, more in line with the typical densities of molecular clouds with active star formation \citep{Hopkins2011, Agertz2013, Dutton2019}. Our study points out that baryonic treatments that are able to successfully reproduce many of the properties of galaxy populations, might not be as well attuned to perform well in environments where the typical conditions of the gas differs from that in the ISM of galaxies, for example in the ICM. As shown in Fig.~\ref{fig:SFR_highz}, the potential limitations of these treatments might worsen at intermediate to high-redshifts, where the associated higher densities translate into larger numbers of gas cells being eligible to become star-forming in prescriptions based on a low density threshold for star formation. In other words, while baryonic treatments such as the one employed in the TNG model, do predict the presence of overdense pockets of gas within the ICM, it is insufficient to uniquely determine which fraction of that gas is dense and cold enough to form stars, a limitation that may scale with redshift as gas is more abundant and dense at high-z. As more data become available for low- and high-z groups and clusters, their imaging in wavelengths where young stars are mapped might help understand the applicability of such star formation recipes to these environments.

\section{Summary and Conclusions}\label{conclusions}

We use the cosmological hydrodynamical simulation TNG50 to study the formation of in-situ ICL in Virgo-like systems. We select the 3 most massive clusters in the box with virial masses $M_{200} = 6.5 \times 10^{13}$-$1.8 \times 10^{14}$\msun. We find that a significant fraction, $\sim 8\%$-$28\%$ of the present-day mass in the ICL, was formed in-situ. When analyzed in detail, and contrary to what is commonly believed, the majority of these in-situ stars are born directly into the diffuse light component at hundreds of kiloparsecs from the main host and directly from the ICM gas, rather than brought in and dispersed during merger events. This diffuse star-formation proceeds in cloudlets that condense along filamentary-like structures that are aligned with the neutral gas and dark matter distributions and are not clearly  associated with ram-pressure stripping of gas in infalling galaxies. 

At $z=0$ the expected H$_\alpha$ surface brightness due to this diffuse mode of star formation reaches $\sim 6.8 \times 10^{-19}$ $\rm erg\;\rm s^{-1}\; cm^{-2}\; \rm arcsec^{-2}$, just below the detection limits of the ongoing surveys in Virgo when smoothed on scales of $3$ arcsec. In the simulated clusters, star formation maps at higher redshift  suggest that extended star formation is present throughout the  history of the cluster, making it an interesting target for discovery with facilities such as JWST. 

One interesting application of our results is that orphan core-collapse supernova events would be expected to occur in relative isolation in the ICL. While such events have not yet been reported, with a predicted average SFR of $\sim 1$ $\mathrm{M}_\odot$/yr, we expect only $\sim$ 1 SN per $\sim 200$ yr in a Virgo-like system\textsuperscript{4}\footnote{We have assumed a power-law mass function following \cite{Salpeter1955}, for the mass range of $0.1$--$20$ \msun\; with stars more massive than $8$ \msun exploding as SN, which results in $\sim 7.2\%$ of the total stellar mass formed able to generate a SN event. Assuming a mean SN mass $12$ \msun, we anticipate the occurrence of 1 SN in $167$ ($\sim 200$) years.}, suggesting that a concurrent survey of hundreds of clusters might be needed for detecting such rare events.

If ongoing star formation in the intra-cluster region of observed clusters is really lacking, then this would signal shortcomings in the treatment of star formation in simulations like TNG50, especially in low-density regions. The formation of stars in these extended and diffuse structures is  possible because of the condensation of denser clouds embedded in the more diffuse ICM, as reported by \citet{Nelson2020}. Numerical techniques such as that used in {\sc Arepo} have been shown to lead to less artificial formation of clouds compared to other techniques such as SPH \citep{Agertz2009, Torrey2012}, but it is possible that other numerical effects are still at play. We emphasize that further studies of the effect of the star formation threshold in models such as those of TNG would be desirable to determine their role in the formation of stars in such unexpected conditions. 

\section*{Acknowledgements}

We gratefully acknowledge the contributions of several individuals who supported this research. Special thanks to our anonymous referee whose valuable suggestions greatly improved this work. The authors would also like to thank Eric Rohr, Annalisa Pillepich, Dylan Nelson, Lars Hernquist and Chris Mihos for insightful discussions that helped shape this paper.  NA and LVS would like to acknowledge the financial support received from the NASA ATP-80NSSC20K0566,  NSF-CAREER-1945310 and NSF-AST-2107993 grants. NA would like to acknowledge the support provided by the UCR-Carnegie Graduate Fellowship.

%%%%%%%%%%%%%%%%%%%%%%%%%%%%%%%%%%%%%%%%%%%%%%%%%%
\section*{Data Availability}

This paper is based on halo catalogs and merger trees from the IllustrisTNG Project \citep{Nelson2019}. These data are publicly available at \href{https://www.tng-project.org/}{https://www.tng-project.org/}.

\newpage

\bibliographystyle{mnras}
\bibliography{manuscript}

\begin{thebibliography}{}
\makeatletter
\relax
\def\mn@urlcharsother{\let\do\@makeother \do\$\do\&\do\#\do\^\do\_\do\%\do\~}
\def\mn@doi{\begingroup\mn@urlcharsother \@ifnextchar [ {\mn@doi@} {\mn@doi@[]}}
\def\mn@doi@[#1]#2{\def\@tempa{#1}\ifx\@tempa\@empty \href {http://dx.doi.org/#2} {doi:#2}\else \href {http://dx.doi.org/#2} {#1}\fi \endgroup}
\def\mn@eprint#1#2{\mn@eprint@#1:#2::\@nil}
\def\mn@eprint@arXiv#1{\href {http://arxiv.org/abs/#1} {{\tt arXiv:#1}}}
\def\mn@eprint@dblp#1{\href {http://dblp.uni-trier.de/rec/bibtex/#1.xml} {dblp:#1}}
\def\mn@eprint@#1:#2:#3:#4\@nil{\def\@tempa {#1}\def\@tempb {#2}\def\@tempc {#3}\ifx \@tempc \@empty \let \@tempc \@tempb \let \@tempb \@tempa \fi \ifx \@tempb \@empty \def\@tempb {arXiv}\fi \@ifundefined {mn@eprint@\@tempb}{\@tempb:\@tempc}{\expandafter \expandafter \csname mn@eprint@\@tempb\endcsname \expandafter{\@tempc}}}

\bibitem[\protect\citeauthoryear{{Afruni}, {Fraternali}  \& {Pezzulli}}{{Afruni} et~al.}{2019}]{Afruni2019}
{Afruni} A.,  {Fraternali} F.,   {Pezzulli} G.,  2019, \mn@doi [\aap] {10.1051/0004-6361/201835002}, \href {https://ui.adsabs.harvard.edu/abs/2019A&A...625A..11A} {625, A11}

\bibitem[\protect\citeauthoryear{{Agertz}, {Teyssier}  \& {Moore}}{{Agertz} et~al.}{2009}]{Agertz2009}
{Agertz} O.,  {Teyssier} R.,   {Moore} B.,  2009, \mn@doi [\mnras] {10.1111/j.1745-3933.2009.00685.x}, \href {https://ui.adsabs.harvard.edu/abs/2009MNRAS.397L..64A} {397, L64}

\bibitem[\protect\citeauthoryear{{Agertz}, {Kravtsov}, {Leitner}  \& {Gnedin}}{{Agertz} et~al.}{2013}]{Agertz2013}
{Agertz} O.,  {Kravtsov} A.~V.,  {Leitner} S.~N.,   {Gnedin} N.~Y.,  2013, \mn@doi [\apj] {10.1088/0004-637X/770/1/25}, \href {https://ui.adsabs.harvard.edu/abs/2013ApJ...770...25A} {770, 25}

\bibitem[\protect\citeauthoryear{{Ahvazi}, {Sales}, {Doppel}, {Benson}, {D'Souza}  \& {Rodriguez-Gomez}}{{Ahvazi} et~al.}{2024}]{Ahvazi2023}
{Ahvazi} N.,  {Sales} L.~V.,  {Doppel} J.~E.,  {Benson} A.,  {D'Souza} R.,   {Rodriguez-Gomez} V.,  2024, \mn@doi [\mnras] {10.1093/mnras/stae848}, \href {https://ui.adsabs.harvard.edu/abs/2024MNRAS.529.4666A} {529, 4666}

\bibitem[\protect\citeauthoryear{{Barfety} et~al.,}{{Barfety} et~al.}{2022}]{Barfety2022}
{Barfety} C.,  et~al., 2022, \mn@doi [\apj] {10.3847/1538-4357/ac61dd}, \href {https://ui.adsabs.harvard.edu/abs/2022ApJ...930...25B} {930, 25}

\bibitem[\protect\citeauthoryear{Battaia et~al.,}{Battaia et~al.}{2012}]{ArrigoniBattaia2012}
Battaia F.~A.,  et~al., 2012, \mn@doi [Astronomy & Astrophysics] {10.1051/0004-6361/201218895}, 543, A112

\bibitem[\protect\citeauthoryear{{Beccari} et~al.,}{{Beccari} et~al.}{2017}]{Beccari2017}
{Beccari} G.,  et~al., 2017, \mn@doi [\mnras] {10.1093/mnras/stw2874}, \href {https://ui.adsabs.harvard.edu/abs/2017MNRAS.465.2189B} {465, 2189}

\bibitem[\protect\citeauthoryear{{Bellazzini} et~al.,}{{Bellazzini} et~al.}{2018}]{Bellazzini2018}
{Bellazzini} M.,  et~al., 2018, \mn@doi [\mnras] {10.1093/mnras/sty467}, \href {https://ui.adsabs.harvard.edu/abs/2018MNRAS.476.4565B} {476, 4565}

\bibitem[\protect\citeauthoryear{{Boselli}, {Fossati}, {Gavazzi}, {Ciesla}, {Buat}, {Boissier}  \& {Hughes}}{{Boselli} et~al.}{2015}]{Boselli2015}
{Boselli} A.,  {Fossati} M.,  {Gavazzi} G.,  {Ciesla} L.,  {Buat} V.,  {Boissier} S.,   {Hughes} T.~M.,  2015, \mn@doi [\aap] {10.1051/0004-6361/201525712}, \href {https://ui.adsabs.harvard.edu/abs/2015A&A...579A.102B} {579, A102}

\bibitem[\protect\citeauthoryear{{Boselli} et~al.,}{{Boselli} et~al.}{2018a}]{Boselli2018}
{Boselli} A.,  et~al., 2018a, \mn@doi [\aap] {10.1051/0004-6361/201732407}, \href {https://ui.adsabs.harvard.edu/abs/2018A&A...614A..56B} {614, A56}

\bibitem[\protect\citeauthoryear{{Boselli} et~al.,}{{Boselli} et~al.}{2018b}]{Boselli2018VESTIGEIII}
{Boselli} A.,  et~al., 2018b, \mn@doi [\aap] {10.1051/0004-6361/201732410}, \href {https://ui.adsabs.harvard.edu/abs/2018A&A...615A.114B} {615, A114}

\bibitem[\protect\citeauthoryear{{Boselli}, {Fossati}  \& {Sun}}{{Boselli} et~al.}{2022}]{Boselli2022}
{Boselli} A.,  {Fossati} M.,   {Sun} M.,  2022, \mn@doi [\aapr] {10.1007/s00159-022-00140-3}, \href {https://ui.adsabs.harvard.edu/abs/2022A&ARv..30....3B} {30, 3}

\bibitem[\protect\citeauthoryear{{Boselli} et~al.,}{{Boselli} et~al.}{2023}]{Boselli2023}
{Boselli} A.,  et~al., 2023, \mn@doi [\aap] {10.1051/0004-6361/202244267}, \href {https://ui.adsabs.harvard.edu/abs/2023A&A...669A..73B} {669, A73}

\bibitem[\protect\citeauthoryear{{Bundy} et~al.,}{{Bundy} et~al.}{2015}]{Bundy2015}
{Bundy} K.,  et~al., 2015, \mn@doi [\apj] {10.1088/0004-637X/798/1/7}, \href {https://ui.adsabs.harvard.edu/abs/2015ApJ...798....7B} {798, 7}

\bibitem[\protect\citeauthoryear{{Butsky}, {Burchett}, {Nagai}, {Tremmel}, {Quinn}  \& {Werk}}{{Butsky} et~al.}{2019}]{Butsky2019}
{Butsky} I.~S.,  {Burchett} J.~N.,  {Nagai} D.,  {Tremmel} M.,  {Quinn} T.~R.,   {Werk} J.~K.,  2019, \mn@doi [\mnras] {10.1093/mnras/stz2859}, \href {https://ui.adsabs.harvard.edu/abs/2019MNRAS.490.4292B} {490, 4292}

\bibitem[\protect\citeauthoryear{{Calura}, {Bellazzini}  \& {D'Ercole}}{{Calura} et~al.}{2020}]{Calura2020}
{Calura} F.,  {Bellazzini} M.,   {D'Ercole} A.,  2020, \mn@doi [\mnras] {10.1093/mnras/staa3133}, \href {https://ui.adsabs.harvard.edu/abs/2020MNRAS.499.5873C} {499, 5873}

\bibitem[\protect\citeauthoryear{{Chabrier}}{{Chabrier}}{2003}]{Chabrier2003}
{Chabrier} G.,  2003, \mn@doi [\apjl] {10.1086/374879}, \href {https://ui.adsabs.harvard.edu/abs/2003ApJ...586L.133C} {586, L133}

\bibitem[\protect\citeauthoryear{Contini, {De Lucia}, Villalobos  \& Borgani}{Contini et~al.}{2013}]{Contini2013}
Contini E.,  {De Lucia} G.,  Villalobos A.,   Borgani S.,  2013, \mn@doi [The Astrophysical Journal] {10.1093/mnras/stt2174}, 699, 1518

\bibitem[\protect\citeauthoryear{Contini, Yi  \& Kang}{Contini et~al.}{2018}]{Contini2018}
Contini E.,  Yi S.~K.,   Kang X.,  2018, \mn@doi [Monthly Notices of the Royal Astronomical Society] {10.1093/mnras/sty1518}, 592, A7

\bibitem[\protect\citeauthoryear{{Contini}, {Yi}  \& {Kang}}{{Contini} et~al.}{2019}]{Contini2019}
{Contini} E.,  {Yi} S.~K.,   {Kang} X.,  2019, \mn@doi [\apj] {10.3847/1538-4357/aaf41f}, \href {https://ui.adsabs.harvard.edu/abs/2019ApJ...871...24C} {871, 24}

\bibitem[\protect\citeauthoryear{Cooper, D'Souza, Kauffmann, Wang, Boylan-Kolchin, Guo, Frenk  \& White}{Cooper et~al.}{2013}]{Cooper2013}
Cooper A.~P.,  D'Souza R.,  Kauffmann G.,  Wang J.,  Boylan-Kolchin M.,  Guo Q.,  Frenk C.~S.,   White S. D.~M.,  2013, \mn@doi [Monthly Notices of the Royal Astronomical Society] {10.1093/mnras/stt1245}, 434, 3348

\bibitem[\protect\citeauthoryear{Cooper, Gao, Guo, Frenk, Jenkins, Springel  \& White}{Cooper et~al.}{2015a}]{Cooper2015}
Cooper A.~P.,  Gao L.,  Guo Q.,  Frenk C.~S.,  Jenkins A.,  Springel V.,   White S. D.~M.,  2015a, \mn@doi [Monthly Notices of the Royal Astronomical Society] {10.1093/mnras/stv1042}, 451, 2703

\bibitem[\protect\citeauthoryear{{Cooper}, {Parry}, {Lowing}, {Cole}  \& {Frenk}}{{Cooper} et~al.}{2015b}]{Cooper2015_2}
{Cooper} A.~P.,  {Parry} O.~H.,  {Lowing} B.,  {Cole} S.,   {Frenk} C.,  2015b, \mn@doi [\mnras] {10.1093/mnras/stv2057}, \href {https://ui.adsabs.harvard.edu/abs/2015MNRAS.454.3185C} {454, 3185}

\bibitem[\protect\citeauthoryear{Cui et~al.,}{Cui et~al.}{2014}]{Cui2014}
Cui W.,  et~al., 2014, \mn@doi [Monthly Notices of the Royal Astronomical Society] {10.1093/mnras/stt1940}, 437, 816

\bibitem[\protect\citeauthoryear{{Davis}, {Efstathiou}, {Frenk}  \& {White}}{{Davis} et~al.}{1985}]{1985ApJ...292..371D}
{Davis} M.,  {Efstathiou} G.,  {Frenk} C.~S.,   {White} S.~D.~M.,  1985, \mn@doi [\apj] {10.1086/163168}, \href {https://ui.adsabs.harvard.edu/abs/1985ApJ...292..371D} {292, 371}

\bibitem[\protect\citeauthoryear{{Dolag}, {Borgani}, {Murante}  \& {Springel}}{{Dolag} et~al.}{2009}]{2009MNRAS.399..497D}
{Dolag} K.,  {Borgani} S.,  {Murante} G.,   {Springel} V.,  2009, \mn@doi [\mnras] {10.1111/j.1365-2966.2009.15034.x}, \href {https://ui.adsabs.harvard.edu/abs/2009MNRAS.399..497D} {399, 497}

\bibitem[\protect\citeauthoryear{{Donnari}, {Pillepich}, {Nelson}, {Marinacci}, {Vogelsberger}  \& {Hernquist}}{{Donnari} et~al.}{2021}]{Donnari2021b}
{Donnari} M.,  {Pillepich} A.,  {Nelson} D.,  {Marinacci} F.,  {Vogelsberger} M.,   {Hernquist} L.,  2021, \mn@doi [\mnras] {10.1093/mnras/stab1950}, \href {https://ui.adsabs.harvard.edu/abs/2021MNRAS.506.4760D} {506, 4760}

\bibitem[\protect\citeauthoryear{{Dutton}, {Macci{\`o}}, {Buck}, {Dixon}, {Blank}  \& {Obreja}}{{Dutton} et~al.}{2019}]{Dutton2019}
{Dutton} A.~A.,  {Macci{\`o}} A.~V.,  {Buck} T.,  {Dixon} K.~L.,  {Blank} M.,   {Obreja} A.,  2019, \mn@doi [\mnras] {10.1093/mnras/stz889}, \href {https://ui.adsabs.harvard.edu/abs/2019MNRAS.486..655D} {486, 655}

\bibitem[\protect\citeauthoryear{{Fielding}, {Ostriker}, {Bryan}  \& {Jermyn}}{{Fielding} et~al.}{2020}]{Fielding2020}
{Fielding} D.~B.,  {Ostriker} E.~C.,  {Bryan} G.~L.,   {Jermyn} A.~S.,  2020, \mn@doi [\apjl] {10.3847/2041-8213/ab8d2c}, \href {https://ui.adsabs.harvard.edu/abs/2020ApJ...894L..24F} {894, L24}

\bibitem[\protect\citeauthoryear{{Font}, {McCarthy}, {Crain}, {Theuns}, {Schaye}, {Wiersma}  \& {Dalla Vecchia}}{{Font} et~al.}{2011}]{Font2011}
{Font} A.~S.,  {McCarthy} I.~G.,  {Crain} R.~A.,  {Theuns} T.,  {Schaye} J.,  {Wiersma} R.~P.~C.,   {Dalla Vecchia} C.,  2011, \mn@doi [\mnras] {10.1111/j.1365-2966.2011.19227.x}, \href {https://ui.adsabs.harvard.edu/abs/2011MNRAS.416.2802F} {416, 2802}

\bibitem[\protect\citeauthoryear{{Ford}, {Oppenheimer}, {Dav{\'e}}, {Katz}, {Kollmeier}  \& {Weinberg}}{{Ford} et~al.}{2013}]{Ford2013}
{Ford} A.~B.,  {Oppenheimer} B.~D.,  {Dav{\'e}} R.,  {Katz} N.,  {Kollmeier} J.~A.,   {Weinberg} D.~H.,  2013, \mn@doi [\mnras] {10.1093/mnras/stt393}, \href {https://ui.adsabs.harvard.edu/abs/2013MNRAS.432...89F} {432, 89}

\bibitem[\protect\citeauthoryear{{Fumagalli}, {Gavazzi}, {Scaramella}  \& {Franzetti}}{{Fumagalli} et~al.}{2011}]{Fumagalli2011}
{Fumagalli} M.,  {Gavazzi} G.,  {Scaramella} R.,   {Franzetti} P.,  2011, \mn@doi [\aap] {10.1051/0004-6361/201015463}, \href {https://ui.adsabs.harvard.edu/abs/2011A&A...528A..46F} {528, A46}

\bibitem[\protect\citeauthoryear{{Genel} et~al.,}{{Genel} et~al.}{2014}]{2014MNRAS.445..175G}
{Genel} S.,  et~al., 2014, \mn@doi [\mnras] {10.1093/mnras/stu1654}, \href {https://ui.adsabs.harvard.edu/abs/2014MNRAS.445..175G} {445, 175}

\bibitem[\protect\citeauthoryear{{Genel} et~al.,}{{Genel} et~al.}{2018}]{Genel2018}
{Genel} S.,  et~al., 2018, \mn@doi [\mnras] {10.1093/mnras/stx3078}, \href {https://ui.adsabs.harvard.edu/abs/2018MNRAS.474.3976G} {474, 3976}

\bibitem[\protect\citeauthoryear{{Gerhard}, {Arnaboldi}, {Freeman}  \& {Okamura}}{{Gerhard} et~al.}{2002}]{Gerhard2002}
{Gerhard} O.,  {Arnaboldi} M.,  {Freeman} K.~C.,   {Okamura} S.,  2002, \mn@doi [\apjl] {10.1086/345657}, \href {https://ui.adsabs.harvard.edu/abs/2002ApJ...580L.121G} {580, L121}

\bibitem[\protect\citeauthoryear{{G{\"o}ller}, {Joshi}, {Rohr}, {Zinger}  \& {Pillepich}}{{G{\"o}ller} et~al.}{2023}]{Goller2023}
{G{\"o}ller} J.,  {Joshi} G.~D.,  {Rohr} E.,  {Zinger} E.,   {Pillepich} A.,  2023, \mn@doi [\mnras] {10.1093/mnras/stad2551}, \href {https://ui.adsabs.harvard.edu/abs/2023MNRAS.525.3551G} {525, 3551}

\bibitem[\protect\citeauthoryear{{Gullieuszik} et~al.,}{{Gullieuszik} et~al.}{2020}]{Gullieuszik2020}
{Gullieuszik} M.,  et~al., 2020, \mn@doi [\apj] {10.3847/1538-4357/aba3cb}, \href {https://ui.adsabs.harvard.edu/abs/2020ApJ...899...13G} {899, 13}

\bibitem[\protect\citeauthoryear{{Hao}, {Kennicutt}, {Johnson}, {Calzetti}, {Dale}  \& {Moustakas}}{{Hao} et~al.}{2011}]{2011ApJ...741..124H}
{Hao} C.-N.,  {Kennicutt} R.~C.,  {Johnson} B.~D.,  {Calzetti} D.,  {Dale} D.~A.,   {Moustakas} J.,  2011, \mn@doi [\apj] {10.1088/0004-637X/741/2/124}, \href {https://ui.adsabs.harvard.edu/abs/2011ApJ...741..124H} {741, 124}

\bibitem[\protect\citeauthoryear{{Hester} et~al.,}{{Hester} et~al.}{2010}]{Hester2010}
{Hester} J.~A.,  et~al., 2010, \mn@doi [\apjl] {10.1088/2041-8205/716/1/L14}, \href {https://ui.adsabs.harvard.edu/abs/2010ApJ...716L..14H} {716, L14}

\bibitem[\protect\citeauthoryear{{Hlavacek-Larrondo} et~al.,}{{Hlavacek-Larrondo} et~al.}{2020}]{Hlavacek2020}
{Hlavacek-Larrondo} J.,  et~al., 2020, \mn@doi [\apjl] {10.3847/2041-8213/ab9ca5}, \href {https://ui.adsabs.harvard.edu/abs/2020ApJ...898L..50H} {898, L50}

\bibitem[\protect\citeauthoryear{{Hopkins}, {Quataert}  \& {Murray}}{{Hopkins} et~al.}{2011}]{Hopkins2011}
{Hopkins} P.~F.,  {Quataert} E.,   {Murray} N.,  2011, \mn@doi [\mnras] {10.1111/j.1365-2966.2011.19306.x}, \href {https://ui.adsabs.harvard.edu/abs/2011MNRAS.417..950H} {417, 950}

\bibitem[\protect\citeauthoryear{{Jones} et~al.,}{{Jones} et~al.}{2022}]{Jones2022}
{Jones} M.~G.,  et~al., 2022, \mn@doi [\apj] {10.3847/1538-4357/ac7c6c}, \href {https://ui.adsabs.harvard.edu/abs/2022ApJ...935...51J} {935, 51}

\bibitem[\protect\citeauthoryear{{Jones} et~al.,}{{Jones} et~al.}{2024}]{Jones2024}
{Jones} M.~G.,  et~al., 2024, \mn@doi [\apjl] {10.3847/2041-8213/ad3ef5}, \href {https://ui.adsabs.harvard.edu/abs/2024ApJ...966L..15J} {966, L15}

\bibitem[\protect\citeauthoryear{{Junais} et~al.,}{{Junais} et~al.}{2021}]{Junais2021}
{Junais} et~al., 2021, \mn@doi [\aap] {10.1051/0004-6361/202040185}, \href {https://ui.adsabs.harvard.edu/abs/2021A&A...650A..99J} {650, A99}

\bibitem[\protect\citeauthoryear{{Kapferer}, {Sluka}, {Schindler}, {Ferrari}  \& {Ziegler}}{{Kapferer} et~al.}{2009}]{Kapferer2009}
{Kapferer} W.,  {Sluka} C.,  {Schindler} S.,  {Ferrari} C.,   {Ziegler} B.,  2009, \mn@doi [\aap] {10.1051/0004-6361/200811551}, \href {https://ui.adsabs.harvard.edu/abs/2009A&A...499...87K} {499, 87}

\bibitem[\protect\citeauthoryear{{Kaufmann}, {Mayer}, {Wadsley}, {Stadel}  \& {Moore}}{{Kaufmann} et~al.}{2006}]{Kaufmann2006}
{Kaufmann} T.,  {Mayer} L.,  {Wadsley} J.,  {Stadel} J.,   {Moore} B.,  2006, \mn@doi [\mnras] {10.1111/j.1365-2966.2006.10599.x}, \href {https://ui.adsabs.harvard.edu/abs/2006MNRAS.370.1612K} {370, 1612}

\bibitem[\protect\citeauthoryear{{Kaufmann}, {Bullock}, {Maller}, {Fang}  \& {Wadsley}}{{Kaufmann} et~al.}{2009}]{Kaufmann2009}
{Kaufmann} T.,  {Bullock} J.~S.,  {Maller} A.~H.,  {Fang} T.,   {Wadsley} J.,  2009, \mn@doi [\mnras] {10.1111/j.1365-2966.2009.14744.x}, \href {https://ui.adsabs.harvard.edu/abs/2009MNRAS.396..191K} {396, 191}

\bibitem[\protect\citeauthoryear{{Kenney}, {Geha}, {J{\'a}chym}, {Crowl}, {Dague}, {Chung}, {van Gorkom}  \& {Vollmer}}{{Kenney} et~al.}{2014}]{Kenney2014}
{Kenney} J. D.~P.,  {Geha} M.,  {J{\'a}chym} P.,  {Crowl} H.~H.,  {Dague} W.,  {Chung} A.,  {van Gorkom} J.,   {Vollmer} B.,  2014, \mn@doi [\apj] {10.1088/0004-637X/780/2/119}, \href {https://ui.adsabs.harvard.edu/abs/2014ApJ...780..119K} {780, 119}

\bibitem[\protect\citeauthoryear{Kennicutt \& Evans}{Kennicutt \& Evans}{2012}]{2012}
Kennicutt R.~C.,  Evans N.~J.,  2012, \mn@doi [Annual Review of Astronomy and Astrophysics] {10.1146/annurev-astro-081811-125610}, 50, 531–608

\bibitem[\protect\citeauthoryear{{Kronberger}, {Kapferer}, {Unterguggenberger}, {Schindler}  \& {Ziegler}}{{Kronberger} et~al.}{2008}]{Kronberger2008}
{Kronberger} T.,  {Kapferer} W.,  {Unterguggenberger} S.,  {Schindler} S.,   {Ziegler} B.~L.,  2008, \mn@doi [\aap] {10.1051/0004-6361:200809387}, \href {https://ui.adsabs.harvard.edu/abs/2008A&A...483..783K} {483, 783}

\bibitem[\protect\citeauthoryear{{Kroupa}}{{Kroupa}}{2001}]{Kroupa2001}
{Kroupa} P.,  2001, \mn@doi [\mnras] {10.1046/j.1365-8711.2001.04022.x}, \href {https://ui.adsabs.harvard.edu/abs/2001MNRAS.322..231K} {322, 231}

\bibitem[\protect\citeauthoryear{{Lee}, {Kimm}, {Katz}, {Rosdahl}, {Devriendt}  \& {Slyz}}{{Lee} et~al.}{2020}]{Lee2020}
{Lee} J.,  {Kimm} T.,  {Katz} H.,  {Rosdahl} J.,  {Devriendt} J.,   {Slyz} A.,  2020, \mn@doi [\apj] {10.3847/1538-4357/abc3b8}, \href {https://ui.adsabs.harvard.edu/abs/2020ApJ...905...31L} {905, 31}

\bibitem[\protect\citeauthoryear{{Maller} \& {Bullock}}{{Maller} \& {Bullock}}{2004}]{Maller2004}
{Maller} A.~H.,  {Bullock} J.~S.,  2004, \mn@doi [\mnras] {10.1111/j.1365-2966.2004.08349.x}, \href {https://ui.adsabs.harvard.edu/abs/2004MNRAS.355..694M} {355, 694}

\bibitem[\protect\citeauthoryear{{Marinacci} et~al.,}{{Marinacci} et~al.}{2018}]{2018MNRAS.480.5113M}
{Marinacci} F.,  et~al., 2018, \mn@doi [\mnras] {10.1093/mnras/sty2206}, \href {https://ui.adsabs.harvard.edu/abs/2018MNRAS.480.5113M} {480, 5113}

\bibitem[\protect\citeauthoryear{{McCourt}, {Sharma}, {Quataert}  \& {Parrish}}{{McCourt} et~al.}{2012}]{McCourt2012}
{McCourt} M.,  {Sharma} P.,  {Quataert} E.,   {Parrish} I.~J.,  2012, \mn@doi [\mnras] {10.1111/j.1365-2966.2011.19972.x}, \href {https://ui.adsabs.harvard.edu/abs/2012MNRAS.419.3319M} {419, 3319}

\bibitem[\protect\citeauthoryear{Mihos, Harding, Feldmeier, Rudick, Janowiecki, Morrison, Slater  \& Watkins}{Mihos et~al.}{2016}]{Mihos_2016}
Mihos J.~C.,  Harding P.,  Feldmeier J.~J.,  Rudick C.,  Janowiecki S.,  Morrison H.,  Slater C.,   Watkins A.,  2016, \mn@doi [The Astrophysical Journal] {10.3847/1538-4357/834/1/16}, 834, 16

\bibitem[\protect\citeauthoryear{{Montenegro-Taborda}, {Rodriguez-Gomez}, {Pillepich}, {Avila-Reese}, {Sales}, {Rodr{\'\i}guez-Puebla}  \& {Hernquist}}{{Montenegro-Taborda} et~al.}{2023}]{Montenegro2023}
{Montenegro-Taborda} D.,  {Rodriguez-Gomez} V.,  {Pillepich} A.,  {Avila-Reese} V.,  {Sales} L.~V.,  {Rodr{\'\i}guez-Puebla} A.,   {Hernquist} L.,  2023, \mn@doi [\mnras] {10.1093/mnras/stad586}, \href {https://ui.adsabs.harvard.edu/abs/2023MNRAS.521..800M} {521, 800}

\bibitem[\protect\citeauthoryear{Murphy et~al.,}{Murphy et~al.}{2011}]{2011}
Murphy E.~J.,  et~al., 2011, \mn@doi [The Astrophysical Journal] {10.1088/0004-637x/737/2/67}, 737, 67

\bibitem[\protect\citeauthoryear{{Naiman} et~al.,}{{Naiman} et~al.}{2018}]{2018MNRAS.477.1206N}
{Naiman} J.~P.,  et~al., 2018, \mn@doi [\mnras] {10.1093/mnras/sty618}, \href {https://ui.adsabs.harvard.edu/abs/2018MNRAS.477.1206N} {477, 1206}

\bibitem[\protect\citeauthoryear{{Nelson} et~al.,}{{Nelson} et~al.}{2015}]{2015A&C....13...12N}
{Nelson} D.,  et~al., 2015, \mn@doi [Astronomy and Computing] {10.1016/j.ascom.2015.09.003}, \href {https://ui.adsabs.harvard.edu/abs/2015A&C....13...12N} {13, 12}

\bibitem[\protect\citeauthoryear{{Nelson} et~al.,}{{Nelson} et~al.}{2018a}]{Nelson2018_2}
{Nelson} D.,  et~al., 2018a, \mn@doi [\mnras] {10.1093/mnras/stx3040}, \href {https://ui.adsabs.harvard.edu/abs/2018MNRAS.475..624N} {475, 624}

\bibitem[\protect\citeauthoryear{{Nelson} et~al.,}{{Nelson} et~al.}{2018b}]{Nelson2018}
{Nelson} D.,  et~al., 2018b, \mn@doi [\mnras] {10.1093/mnras/sty656}, \href {https://ui.adsabs.harvard.edu/abs/2018MNRAS.477..450N} {477, 450}

\bibitem[\protect\citeauthoryear{{Nelson} et~al.,}{{Nelson} et~al.}{2019a}]{Nelson2019}
{Nelson} D.,  et~al., 2019a, \mn@doi [Computational Astrophysics and Cosmology] {10.1186/s40668-019-0028-x}, \href {https://ui.adsabs.harvard.edu/abs/2019ComAC...6....2N} {6, 2}

\bibitem[\protect\citeauthoryear{{Nelson} et~al.,}{{Nelson} et~al.}{2019b}]{Nelson2019_2}
{Nelson} D.,  et~al., 2019b, \mn@doi [\mnras] {10.1093/mnras/stz2306}, \href {https://ui.adsabs.harvard.edu/abs/2019MNRAS.490.3234N} {490, 3234}

\bibitem[\protect\citeauthoryear{{Nelson} et~al.,}{{Nelson} et~al.}{2020}]{Nelson2020}
{Nelson} D.,  et~al., 2020, \mn@doi [\mnras] {10.1093/mnras/staa2419}, \href {https://ui.adsabs.harvard.edu/abs/2020MNRAS.498.2391N} {498, 2391}

\bibitem[\protect\citeauthoryear{{Nelson}, {Pillepich}, {Ayromlou}, {Lee}, {Lehle}, {Rohr}  \& {Truong}}{{Nelson} et~al.}{2024}]{Nelson2024}
{Nelson} D.,  {Pillepich} A.,  {Ayromlou} M.,  {Lee} W.,  {Lehle} K.,  {Rohr} E.,   {Truong} N.,  2024, \mn@doi [\aap] {10.1051/0004-6361/202348608}, \href {https://ui.adsabs.harvard.edu/abs/2024A&A...686A.157N} {686, A157}

\bibitem[\protect\citeauthoryear{{Oppenheimer}, {Schaye}, {Crain}, {Werk}  \& {Richings}}{{Oppenheimer} et~al.}{2018}]{Oppenheimer2018}
{Oppenheimer} B.~D.,  {Schaye} J.,  {Crain} R.~A.,  {Werk} J.~K.,   {Richings} A.~J.,  2018, \mn@doi [\mnras] {10.1093/mnras/sty2281}, \href {https://ui.adsabs.harvard.edu/abs/2018MNRAS.481..835O} {481, 835}

\bibitem[\protect\citeauthoryear{{Oppenheimer}, {Babul}, {Bah{\'e}}, {Butsky}  \& {McCarthy}}{{Oppenheimer} et~al.}{2021}]{Oppenheimer2021}
{Oppenheimer} B.~D.,  {Babul} A.,  {Bah{\'e}} Y.,  {Butsky} I.~S.,   {McCarthy} I.~G.,  2021, \mn@doi [Universe] {10.3390/universe7070209}, \href {https://ui.adsabs.harvard.edu/abs/2021Univ....7..209O} {7, 209}

\bibitem[\protect\citeauthoryear{{Peng} et~al.,}{{Peng} et~al.}{2010}]{Peng2010}
{Peng} Y.-j.,  et~al., 2010, \mn@doi [\apj] {10.1088/0004-637X/721/1/193}, \href {https://ui.adsabs.harvard.edu/abs/2010ApJ...721..193P} {721, 193}

\bibitem[\protect\citeauthoryear{{Pillepich}, {Madau}  \& {Mayer}}{{Pillepich} et~al.}{2015}]{Pillepich2015}
{Pillepich} A.,  {Madau} P.,   {Mayer} L.,  2015, \mn@doi [\apj] {10.1088/0004-637X/799/2/184}, \href {https://ui.adsabs.harvard.edu/abs/2015ApJ...799..184P} {799, 184}

\bibitem[\protect\citeauthoryear{{Pillepich} et~al.,}{{Pillepich} et~al.}{2018a}]{2018MNRAS.473.4077P}
{Pillepich} A.,  et~al., 2018a, \mn@doi [\mnras] {10.1093/mnras/stx2656}, \href {https://ui.adsabs.harvard.edu/abs/2018MNRAS.473.4077P} {473, 4077}

\bibitem[\protect\citeauthoryear{{Pillepich} et~al.,}{{Pillepich} et~al.}{2018b}]{2018MNRAS.475..648P}
{Pillepich} A.,  et~al., 2018b, \mn@doi [\mnras] {10.1093/mnras/stx3112}, \href {https://ui.adsabs.harvard.edu/abs/2018MNRAS.475..648P} {475, 648}

\bibitem[\protect\citeauthoryear{{Pillepich} et~al.,}{{Pillepich} et~al.}{2019}]{Pillepich2019}
{Pillepich} A.,  et~al., 2019, \mn@doi [\mnras] {10.1093/mnras/stz2338}, \href {https://ui.adsabs.harvard.edu/abs/2019MNRAS.490.3196P} {490, 3196}

\bibitem[\protect\citeauthoryear{{Planck Collaboration} et~al.,}{{Planck Collaboration} et~al.}{2016}]{2016A&A...594A..13P}
{Planck Collaboration} et~al., 2016, \mn@doi [\aap] {10.1051/0004-6361/201525830}, \href {https://ui.adsabs.harvard.edu/abs/2016A&A...594A..13P} {594, A13}

\bibitem[\protect\citeauthoryear{Puchwein, Springel, Sijacki  \& Dolag}{Puchwein et~al.}{2010}]{Puchwein_2010}
Puchwein E.,  Springel V.,  Sijacki D.,   Dolag K.,  2010, \mn@doi [Monthly Notices of the Royal Astronomical Society] {10.1111/j.1365-2966.2010.16786.x}, 447, no

\bibitem[\protect\citeauthoryear{{Rodriguez-Gomez} et~al.,}{{Rodriguez-Gomez} et~al.}{2015}]{2015MNRAS.449...49R}
{Rodriguez-Gomez} V.,  et~al., 2015, \mn@doi [\mnras] {10.1093/mnras/stv264}, \href {https://ui.adsabs.harvard.edu/abs/2015MNRAS.449...49R} {449, 49}

\bibitem[\protect\citeauthoryear{{Rodriguez-Gomez} et~al.,}{{Rodriguez-Gomez} et~al.}{2016}]{Rodriguez2016}
{Rodriguez-Gomez} V.,  et~al., 2016, \mn@doi [\mnras] {10.1093/mnras/stw456}, \href {https://ui.adsabs.harvard.edu/abs/2016MNRAS.458.2371R} {458, 2371}

\bibitem[\protect\citeauthoryear{{Rohr}, {Pillepich}, {Nelson}, {Zinger}, {Joshi}  \& {Ayromlou}}{{Rohr} et~al.}{2023}]{Rohr2023}
{Rohr} E.,  {Pillepich} A.,  {Nelson} D.,  {Zinger} E.,  {Joshi} G.~D.,   {Ayromlou} M.,  2023, \mn@doi [\mnras] {10.1093/mnras/stad2101}, \href {https://ui.adsabs.harvard.edu/abs/2023MNRAS.524.3502R} {524, 3502}

\bibitem[\protect\citeauthoryear{{Rohr}, {Pillepich}, {Nelson}, {Ayromlou}, {P{\'e}roux}  \& {Zinger}}{{Rohr} et~al.}{2024}]{Rohr2024}
{Rohr} E.,  {Pillepich} A.,  {Nelson} D.,  {Ayromlou} M.,  {P{\'e}roux} C.,   {Zinger} E.,  2024, \mn@doi [arXiv e-prints] {10.48550/arXiv.2410.19900}, \href {https://ui.adsabs.harvard.edu/abs/2024arXiv241019900R} {p. arXiv:2410.19900}

\bibitem[\protect\citeauthoryear{{Salpeter}}{{Salpeter}}{1955}]{Salpeter1955}
{Salpeter} E.~E.,  1955, \mn@doi [\apj] {10.1086/145971}, \href {https://ui.adsabs.harvard.edu/abs/1955ApJ...121..161S} {121, 161}

\bibitem[\protect\citeauthoryear{{Sharma}, {McCourt}, {Quataert}  \& {Parrish}}{{Sharma} et~al.}{2012}]{Sharma2012}
{Sharma} P.,  {McCourt} M.,  {Quataert} E.,   {Parrish} I.~J.,  2012, \mn@doi [\mnras] {10.1111/j.1365-2966.2011.20246.x}, \href {https://ui.adsabs.harvard.edu/abs/2012MNRAS.420.3174S} {420, 3174}

\bibitem[\protect\citeauthoryear{{Smith} et~al.,}{{Smith} et~al.}{2010}]{Smith2010}
{Smith} R.~J.,  et~al., 2010, \mn@doi [\mnras] {10.1111/j.1365-2966.2010.17253.x}, \href {https://ui.adsabs.harvard.edu/abs/2010MNRAS.408.1417S} {408, 1417}

\bibitem[\protect\citeauthoryear{{Speagle}, {Steinhardt}, {Capak}  \& {Silverman}}{{Speagle} et~al.}{2014}]{Speagle2014}
{Speagle} J.~S.,  {Steinhardt} C.~L.,  {Capak} P.~L.,   {Silverman} J.~D.,  2014, \mn@doi [\apjs] {10.1088/0067-0049/214/2/15}, \href {https://ui.adsabs.harvard.edu/abs/2014ApJS..214...15S} {214, 15}

\bibitem[\protect\citeauthoryear{{Springel}}{{Springel}}{2010}]{2010MNRAS.401..791S}
{Springel} V.,  2010, \mn@doi [\mnras] {10.1111/j.1365-2966.2009.15715.x}, \href {https://ui.adsabs.harvard.edu/abs/2010MNRAS.401..791S} {401, 791}

\bibitem[\protect\citeauthoryear{{Springel} \& {Hernquist}}{{Springel} \& {Hernquist}}{2003}]{Springel2003}
{Springel} V.,  {Hernquist} L.,  2003, \mn@doi [\mnras] {10.1046/j.1365-8711.2003.06206.x}, \href {https://ui.adsabs.harvard.edu/abs/2003MNRAS.339..289S} {339, 289}

\bibitem[\protect\citeauthoryear{{Springel}, {White}, {Tormen}  \& {Kauffmann}}{{Springel} et~al.}{2001}]{2001MNRAS.328..726S}
{Springel} V.,  {White} S. D.~M.,  {Tormen} G.,   {Kauffmann} G.,  2001, \mn@doi [\mnras] {10.1046/j.1365-8711.2001.04912.x}, \href {https://ui.adsabs.harvard.edu/abs/2001MNRAS.328..726S} {328, 726}

\bibitem[\protect\citeauthoryear{{Springel} et~al.,}{{Springel} et~al.}{2018a}]{2018MNRAS.475..676S}
{Springel} V.,  et~al., 2018a, \mn@doi [\mnras] {10.1093/mnras/stx3304}, \href {https://ui.adsabs.harvard.edu/abs/2018MNRAS.475..676S} {475, 676}

\bibitem[\protect\citeauthoryear{{Springel} et~al.,}{{Springel} et~al.}{2018b}]{Springel2018}
{Springel} V.,  et~al., 2018b, \mn@doi [\mnras] {10.1093/mnras/stx3304}, \href {https://ui.adsabs.harvard.edu/abs/2018MNRAS.475..676S} {475, 676}

\bibitem[\protect\citeauthoryear{{Steyrleithner}, {Hensler}  \& {Boselli}}{{Steyrleithner} et~al.}{2020}]{Steyrleithner2020}
{Steyrleithner} P.,  {Hensler} G.,   {Boselli} A.,  2020, \mn@doi [\mnras] {10.1093/mnras/staa775}, \href {https://ui.adsabs.harvard.edu/abs/2020MNRAS.494.1114S} {494, 1114}

\bibitem[\protect\citeauthoryear{{Tacchella} et~al.,}{{Tacchella} et~al.}{2022}]{Tacchella2022}
{Tacchella} S.,  et~al., 2022, \mn@doi [\mnras] {10.1093/mnras/stac818}, \href {https://ui.adsabs.harvard.edu/abs/2022MNRAS.513.2904T} {513, 2904}

\bibitem[\protect\citeauthoryear{{Tissera}, {White}  \& {Scannapieco}}{{Tissera} et~al.}{2012}]{Tissera2012}
{Tissera} P.~B.,  {White} S. D.~M.,   {Scannapieco} C.,  2012, \mn@doi [\mnras] {10.1111/j.1365-2966.2011.20028.x}, \href {https://ui.adsabs.harvard.edu/abs/2012MNRAS.420..255T} {420, 255}

\bibitem[\protect\citeauthoryear{{Tonnesen} \& {Bryan}}{{Tonnesen} \& {Bryan}}{2012}]{Tonnesen2012}
{Tonnesen} S.,  {Bryan} G.~L.,  2012, \mn@doi [\mnras] {10.1111/j.1365-2966.2012.20737.x}, \href {https://ui.adsabs.harvard.edu/abs/2012MNRAS.422.1609T} {422, 1609}

\bibitem[\protect\citeauthoryear{{Torrey}, {Vogelsberger}, {Sijacki}, {Springel}  \& {Hernquist}}{{Torrey} et~al.}{2012}]{Torrey2012}
{Torrey} P.,  {Vogelsberger} M.,  {Sijacki} D.,  {Springel} V.,   {Hernquist} L.,  2012, \mn@doi [\mnras] {10.1111/j.1365-2966.2012.22082.x}, \href {https://ui.adsabs.harvard.edu/abs/2012MNRAS.427.2224T} {427, 2224}

\bibitem[\protect\citeauthoryear{{Vogelsberger}, {Genel}, {Sijacki}, {Torrey}, {Springel}  \& {Hernquist}}{{Vogelsberger} et~al.}{2013a}]{2013MNRAS.436.3031V}
{Vogelsberger} M.,  {Genel} S.,  {Sijacki} D.,  {Torrey} P.,  {Springel} V.,   {Hernquist} L.,  2013a, \mn@doi [\mnras] {10.1093/mnras/stt1789}, \href {https://ui.adsabs.harvard.edu/abs/2013MNRAS.436.3031V} {436, 3031}

\bibitem[\protect\citeauthoryear{{Vogelsberger}, {Genel}, {Sijacki}, {Torrey}, {Springel}  \& {Hernquist}}{{Vogelsberger} et~al.}{2013b}]{Vogelsberger2013}
{Vogelsberger} M.,  {Genel} S.,  {Sijacki} D.,  {Torrey} P.,  {Springel} V.,   {Hernquist} L.,  2013b, \mn@doi [\mnras] {10.1093/mnras/stt1789}, \href {https://ui.adsabs.harvard.edu/abs/2013MNRAS.436.3031V} {436, 3031}

\bibitem[\protect\citeauthoryear{{Vogelsberger} et~al.,}{{Vogelsberger} et~al.}{2014a}]{2014MNRAS.444.1518V}
{Vogelsberger} M.,  et~al., 2014a, \mn@doi [\mnras] {10.1093/mnras/stu1536}, \href {https://ui.adsabs.harvard.edu/abs/2014MNRAS.444.1518V} {444, 1518}

\bibitem[\protect\citeauthoryear{{Vogelsberger} et~al.,}{{Vogelsberger} et~al.}{2014b}]{2014Natur.509..177V}
{Vogelsberger} M.,  et~al., 2014b, \mn@doi [\nat] {10.1038/nature13316}, \href {https://ui.adsabs.harvard.edu/abs/2014Natur.509..177V} {509, 177}

\bibitem[\protect\citeauthoryear{{Vollmer}, {Braine}, {Mazzilli-Ciraulo}  \& {Schneider}}{{Vollmer} et~al.}{2021}]{Vollmer2021}
{Vollmer} B.,  {Braine} J.,  {Mazzilli-Ciraulo} B.,   {Schneider} B.,  2021, \mn@doi [\aap] {10.1051/0004-6361/202037887}, \href {https://ui.adsabs.harvard.edu/abs/2021A&A...647A.138V} {647, A138}

\bibitem[\protect\citeauthoryear{{Webb} et~al.,}{{Webb} et~al.}{2015a}]{Webb2015}
{Webb} T.,  et~al., 2015a, \mn@doi [\apj] {10.1088/0004-637X/809/2/173}, \href {https://ui.adsabs.harvard.edu/abs/2015ApJ...809..173W} {809, 173}

\bibitem[\protect\citeauthoryear{{Webb} et~al.,}{{Webb} et~al.}{2015b}]{Webb2015b}
{Webb} T. M.~A.,  et~al., 2015b, \mn@doi [\apj] {10.1088/0004-637X/814/2/96}, \href {https://ui.adsabs.harvard.edu/abs/2015ApJ...814...96W} {814, 96}

\bibitem[\protect\citeauthoryear{{Weinberger} et~al.,}{{Weinberger} et~al.}{2017}]{Weinberger2017}
{Weinberger} R.,  et~al., 2017, \mn@doi [\mnras] {10.1093/mnras/stw2944}, \href {https://ui.adsabs.harvard.edu/abs/2017MNRAS.465.3291W} {465, 3291}

\bibitem[\protect\citeauthoryear{{Zolotov}, {Willman}, {Brooks}, {Governato}, {Brook}, {Hogg}, {Quinn}  \& {Stinson}}{{Zolotov} et~al.}{2009}]{Zolotov2009}
{Zolotov} A.,  {Willman} B.,  {Brooks} A.~M.,  {Governato} F.,  {Brook} C.~B.,  {Hogg} D.~W.,  {Quinn} T.,   {Stinson} G.,  2009, \mn@doi [\apj] {10.1088/0004-637X/702/2/1058}, \href {https://ui.adsabs.harvard.edu/abs/2009ApJ...702.1058Z} {702, 1058}

\makeatother
\end{thebibliography}

%%%%%%%%%%%%%%%%%%%%%%%%%%%%%%%%%%%%%%%%%%%%%%%%%%

%%%%%%%%%%%%%%%%% APPENDICES %%%%%%%%%%%%%%%%%%%%%

\appendix
\section{Physical conditions for star formation} \label{SF_condition}

\begin{figure}
    \center
	\includegraphics[width=0.5\columnwidth]{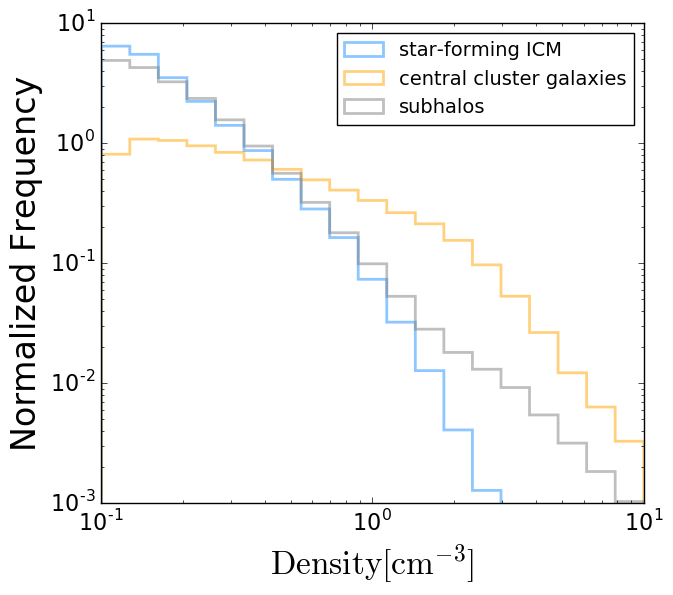}
    \caption{Normalized distribution of star-forming gas cells that belong to the ICM (shown by the blue line), massive subhalos (presented by the gray line), and central cluster galaxies (depicted by the orange line).}
    \label{fig:app2}
\end{figure}

In this section we provide additional information on the gas density of the star-forming gas. We show in Fig.~\ref{fig:app2} the normalized distribution of gas density of star-forming gas in different environments: ICM (blue), satellites(gray) and central cluster galaxies (orange) in our clusters. The classification of cells as either belonging to the ICM or to satellites has been done consistent with the classification of stars, following membership of cells as assigned by {\sc Subfind} and ``cleaning" the ICM of particles close in distance to surviving satellites (see Sec.~\ref{ssec:cleaning}). Fig.~\ref{fig:app2} shows that the star formation in the ICM occurs with densities biased slightly low with respect to the star formation in the ISM of galaxies, but there is considerable overlap in the distributions. 

%%%%%%%%%%%%%%%%%%%%%%%%%%%%%%%%%%%%%%%%%%%%%%%%%%
%%%%%%%%%%%%%%%%%%%%%%%%%%%%%%%%%%%%%%%%%%%%%%%%%%

\section{Effect of resolution} \label{res_effect}

To assess the impact of resolution on the star formation rates found within the ICL regions in our studied groups in TNG50, we turn our attention to the results obtained from TNG100 and TNG50-2. TNG100-1, part of the same suite of cosmological hydrodynamical simulations, represents a larger baryonic run with a volume of $110.7 \, \rm Mpc^3$ and $2 \times 1820^3$ resolution elements (lower resolution compared to TNG50, which features $2 \times 2160^3$ resolution elements). In this run, the baryon and dark matter particle masses are $1.4 \times 10^6$\msun and $7.5 \times 10^6$\msun, respectively. TNG50-2 has the exact same volume as TNG50 but simulated at the resolution of TNG100.

For the comparison, we exclusively consider groups with virial masses similar to those examined in our study ($M_{\rm 200} \gtrapprox 6.5 \times 10^{13}$), resulting in a subset of 26 groups in TNG100 and 3 groups in TNG50-2.

\begin{figure}
    \center
	\includegraphics[width=0.5\columnwidth]{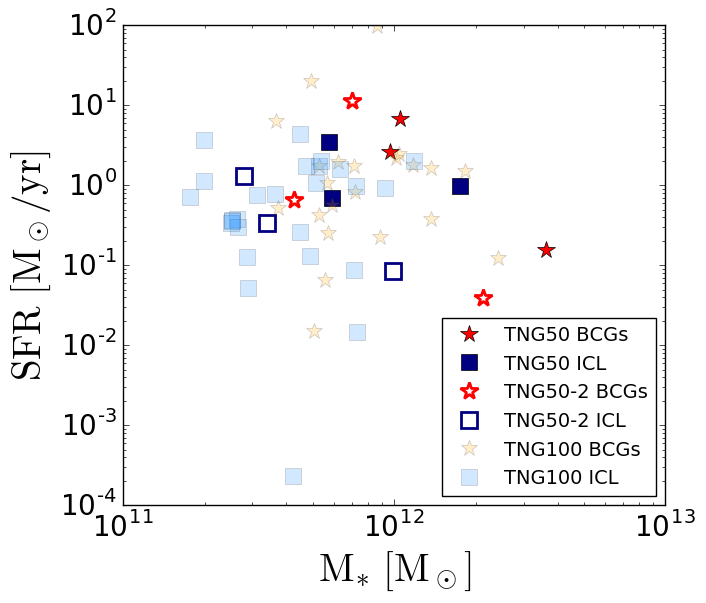}
    \caption{Star formation rate (SFR) plotted against stellar mass for different groups in in TNG100, TNG50, and TNG50-2 simulations. Results for central galaxies are represented by stars, with orange denoting centrals in TNG100, red for TNG50, and unfilled red stars for TNG50-2. The ICL component is illustrated by squares, with light blue indicating ICL in TNG100, navy blue for TNG50, and unfilled navy blue representing TNG50-2.}
    \label{fig:app1}
\end{figure}

In Fig.~\ref{fig:app1}, we juxtapose the star formation rates in TNG50 and TNG100 for both central and ICL regions. 
The results of TNG50 seem within the scatter of those for the larger sample of TNG100 groups, perhaps slightly biased to the upper half of the SFR distribution. For a more direct comparison, we also show the 3 most massive groups in the TNG50-2 run, close in resolution to TNG100. These comparisons suggests a weak dependence of the results with numerical resolution which is in line with the resolution trends found in the formation of clouds in \citet{Nelson2020}: better resolved runs seem more prone to form clouds and also will result in a larger chance to have star-forming gas in them. \\

%%%%%%%%%%%%%%%%%%%%%%%%%%%%%%%%%%%%%%%%%%%%%%%%%%
%%%%%%%%%%%%%%%%%%%%%%%%%%%%%%%%%%%%%%%%%%%%%%%%%%

\end{document}